\documentclass[twocolumn,aps,amsmath,amssymb,prl]{revtex4}
\usepackage{epsfig,bm,dcolumn}
\usepackage[usenames]{color}
\usepackage{graphicx}
\begin{document}

\newcommand {\beq} {\begin{equation}}
\newcommand {\eeq} {\end{equation}}
\newcommand {\bqa} {\begin{eqnarray}}
\newcommand {\eqa} {\end{eqnarray}}
\newcommand {\xx} {\ensuremath{{\bf{x}}}}
\newcommand {\kk} {\ensuremath{{\bf{k}}}}
\newcommand {\kp} {\ensuremath{{\bf{k'}}}}
\newcommand {\ca} {\ensuremath{c^{\dagger}}}
\newcommand {\no} {\nonumber}
\newcommand {\sib} {\ensuremath{\overline{\psi}}}
\newcommand {\br} {\ensuremath{\begin{array}{clrr}}}
\newcommand {\er} {\ensuremath{\end{array}}}
\newcommand {\MM} {\ensuremath{{\bf{M}}}}
\newcommand {\A} {\ensuremath{{\bf{A}}}}
\newcommand {\DD} {\ensuremath{{\bf{D}}}}
\newcommand {\LL} {\ensuremath{{\bf{L}}}}
\newcommand {\UU} {\ensuremath{{\bf{U}}}}
\newcommand {\WW} {\ensuremath{{\bf{W}}}}
\newcommand {\VV} {\ensuremath{{\bf{V}}}}
\newcommand {\dn} {\ensuremath{\downarrow}}
\newcommand {\gr} {\ensuremath{{\bf {G}}}}
\newcommand {\up} {\ensuremath{\uparrow}}
\newcommand {\Ds} {\ensuremath{\Delta^*}}
\newcommand {\dd} {\ensuremath{\Delta}}
\newcommand {\qq} {\ensuremath{{\bf{q}}}}
\newcommand {\QQ} {\ensuremath{{\bf{Q}}}}
\newcommand {\KK} {\ensuremath{{\bf{K}}}}
\newcommand {\dl} {\ensuremath{\Delta_0}}
\newcommand {\ep} {\ensuremath{\epsilon}}
\newcommand {\Do} {\ensuremath{\Delta_0}}
\newcommand {\ms} {\medskip}

\title{Quantum Fluctuations in the Superfluid State of the BCS-BEC Crossover}
\author{Roberto B. Diener} 
\author{Rajdeep Sensarma}
\author{Mohit Randeria} 
\affiliation{Department of Physics, The Ohio State University, Columbus, Ohio 43210}
\begin{abstract}
We determine the effects of quantum fluctuations about the $T=0$ mean field solution
of the BCS-BEC crossover in a dilute Fermi gas using the functional integral method. 
These fluctuations are described in terms of the zero point motion of collective  
modes and the virtual scattering of gapped quasiparticles. We calculate their effects on
various measurable properties, including chemical potential, ground state energy, the gap, the
speed of sound and the Landau critical velocity. 
At unitarity, we find excellent agreement with quantum Monte Carlo and experimental results.  
In the BCS limit, we show analytically that we obtain 
Fermi liquid interaction corrections to thermodynamics including the Hartree shift. 
In the BEC limit, we show that the theory leads to an approximate description of the 
reduction of the  scattering length for bosonic molecules and also obtain quantum depletion of the 
Lee-Yang form.  At the end of the paper, we describe a method to 
include feedback of quantum fluctuations into the gap equation,
and discuss the problems of self-consistent calculations in satisfying
Goldstone's theorem and obtaining ultraviolet finite results at unitarity.     
\end{abstract}
\maketitle 

\section{I. Introduction}

The BCS-BEC crossover \cite{leggett80,eagles,nozieres,BECreview}
is a problem of long standing interest in many-body physics with
implications for a variety of fields including condensed matter, high-energy, nuclear and atomic and
molecular physics. Recent experimental progress in cooling atomic Fermi gases to ultralow
temperatures and tuning the interactions between atoms using the Feshbach resonance
technique has led to an explosion of interest in the BCS-BEC crossover \cite{jin,ketterle,universal1,betaGrimm,betaENS,betaThomas,Randy}.

The theoretical problem is to determine the properties
of a system with two species of fermions, spin-up and down, with equal
masses and densities, interacting via a short range attractive potential
described by a scattering length $a_S$.
The two extremes of the crossover are well understood theoretically. 
Weak attractive interactions characterized by a small, negative scattering
length $a_S$ lead to collective Cooper pairing of atoms and BCS superfluidity.
In the opposite limit of large attraction, characterized by a small, positive scattering 
length $a_S$ one obtains bosonic molecules which exhibit BEC. The intermediate regime, around the
unitary point at which $|a_S| \to \infty$, where one has a strongly interacting
Fermi gas, is the most interesting and least well understood theoretically. 

The original mean field (MF) theory of Leggett \cite{leggett80} and Eagles \cite{eagles} 
does a decent job of describing the entire ${\it T=0}$ crossover at a {\it qualitative} level~\cite{finite T}. 
It has only one additional ingredient to the standard BCS theory: the chemical potential must be determined 
self-consistently along with the pairing gap. This is sufficient to give 
qualitatively reasonable results \cite{engelbrecht97} which evolve smoothly through unitarity all 
the way up to the molecular BEC. 

Recent theoretical and experimental developments have led to a realization
of the quantitative shortcomings of mean field (MF) theory at $T=0$ especially
at unitarity. The ground state energy density at unitarity is of
the form ${\cal E}_0/N = (1+\beta)(3\, \epsilon_f / 5)$, which is a ``universal'' number \cite{universal1,universal2}
times the free Fermi gas energy since there is no scale other than $\epsilon_f$ as $|a_S| \to \infty$.
Quantum Monte Carlo (QMC) calculations \cite{Carlson,Giorgini} obtain $(1 + \beta)=0.44$,
while experiments \cite{betaGrimm,betaENS,betaThomas} find $(1 + \beta)$ in the range $0.32$ to $0.44$.
In contrast the ground state energy density within MF theory \cite{engelbrecht97} 
yields $(1 + \beta)=0.59$, which is about $34\%$ larger than the QMC result.  Furthermore, in the BEC limit ($k_Fa_S \rightarrow 0^+$), although the MFT correctly predicts a repulsive interaction between the constituent bosons, it misses  ${\cal O}(k_Fa_S)^{3/2}$
corrections to thermodynamic quantities , which are present in a weakly repulsive Bose gas.

Our main motivations were to understand this
quantitative discrepancy, on which there has been recent progress 
by several approaches \cite{Hu-Liu,son,Sachdev,Leo,Haussman}
(discussed below), and also to get a physical picture of the quantum fluctuations missing in
MF theory that are responsible for such a large energy difference.
We show here that the many-body ground state in the crossover must include, in addition to
BCS pairing, the effects of the zero point motion of the collective excitations -- the oscillation of the 
phase (and amplitude) of the order parameter -- and the effects of virtual scattering of quasiparticle
excitations. 

Our central result, from which essentially
all our other results follow, is that the thermodynamic potential $\Omega$ at $T=0$ is given by
\begin{eqnarray}
\Omega&=& -\frac{m}{4\pi a_S}\dl^2-\sum_\kk\left(E_\kk-\ep_\kk+\mu
-\frac{1}{2}\frac{\dl^2}{\ep_\kk}\right)
\label{thermodynamic_potential}
\\
&& +\frac{1}{2}\sum_\qq\left[\omega_0(\qq)-E_c(\qq)- \displaystyle\int_{-\infty}^ {-E_c(\qq)}\frac{d\omega}{\pi}\delta(\qq,\omega)\right]
+ {\cal R} 
\no
\end{eqnarray}
Here the first line represents the ``fermionic contribution'' to the ground state energy of the superfluid.
It has the same structure as the mean field result and may be thought of as
coming from filling up the negative energy states in a Bogoliubov deGennes framework.
The second line represents the ``bosonic contribution'' which arises
from Gaussian fluctuations about the saddle point. It consists of the zero-point energy of
the collective mode with dispersion $\omega_0(\qq)$ and of an integral which
describes the contribution from the virtual scattering of quasiparticles (with a phase shift $\delta (\bf{q}, \omega)$), whose two-particle continuum begins at $E_c(\qq)$.
The last term ${\cal R}$ regularizes the ultraviolet divergence in the bosonic contribution and 
will be described in detail later; see eq.~(\ref{en_cont}).

The problem of determining the ground state energy density of a strongly interacting system
is analogous to that of determining the cosmological constant in quantum field theories.
The latter problem is notorious for being dominated by physics at the scale of the ultraviolet cutoff. 
Here we will show that our results are independent of the momentum cutoff, which
is the inverse of the range $r_{\rm eff}$ of the attractive potential between fermions. 

Although there has been considerable attention devoted to ``universality'' at unitarity $(|a_S| = \infty)$, 
we emphasize the simple but often overlooked point that any observable quantity at $T=0$ is a universal 
function of the single parameter $1/k_F a_S$. By universal we mean that the results are independent of microscopic 
details below the ultraviolet cutoff length scale of $r_{\rm eff}$, provided $k_F r_{\rm eff} \ll 1$. 
Thus it does not matter whether one looks at an experiment with $^6$Li or $^{40}$K,  the result can only depend on 
system parameters through the combination $1/k_F a_S$.
In the absence of a  small parameter in the crossover problem, we judge the validity of our
approximations not only by their success at unitarity in comparison with quantum Monte Carlo
and experiments, but also by their ability to reproduce known results in the BCS and BEC limits.

\begin{table}
\begin{center}
\begin{tabular}{|c|c|c|c|}
\hline
 & & &\\
$|a_S|=\infty$ & ${\cal E}/\left(3\ep_f/5\right)$ & $c/v_f$& $\Delta_0$ \\
& & & \\
\hline
 & & & \\
Mean Field & $0.59$ & $0.44$ & $0.69$ \\
Theory~\cite{engelbrecht97} & & &\\
\hline
 & & &\\
 Quantum & $0.44$ & $0.38$ & $0.6\pm 0.1$ \\
Monte Carlo~\cite{Carlson,Giorgini}& & &\\
\hline
 & & &\\
 Experiments & $0.32 - 0.51$ & $0.38$& - \\
 \cite{betaGrimm,betaENS,betaThomas}& & & \\
\hline
 & & & \\
Mean Field $+$& $0.40$ & $0.37$& $0.47$\\
Fluctuations & & & \\
\hline
\end{tabular}
\caption{Comparison of ground state energy, speed of sound and gap at unitarity obtained by different methods.  The last row gives the results obtained in sections V and VIII.  The results of a self-consistent calculation are described in section IX.}
\label{table_unit}
\end{center}
\end{table}

We conclude this Section with a summary of our main results and an outline of the rest of the paper.
In Sections II, III and IV we describe the functional integral formalism used in the paper and results for the
chemical potential, gap, ground state energy, speed of sound and Landau critical velocity
as a function of $1/(k_Fa_S)$ are presented in Section V.   
Our main results include: 

\textbf{(a)} In the extreme BCS limit, the fluctuation corrections are dominated by the virtual
scattering of fermionic quasiparticles. We show in Section VI that we recover the exact Fermi-liquid corrections
to the thermodynamics of a dilute Fermi gas, which are the Hartree shift of order $k_F a_S$ and 
the Galitskii and Huang-Lee-Yang corrections \cite{galitskii,Lee-Yang,Fetter-Walecka} of order $(k_F a_S)^2$, albeit with a negative scattering length $a_S$.
We note that these are obtained from Gaussian fluctuations about the {\em broken symmetry}
state and not by including them in an ad hoc way.

\textbf{(b)} In the extreme BEC limit, the zero-point motion of the Bogoliubov sound mode dominates the
thermodynamics. From the leading order corrections we estimate in Section VII the effective
scattering length between the molecular bosons to be $a_B \simeq 0.55 a_S$, an
approximate result which turns out to be close to the exact result \cite{petrov} for the four-body
problem of $0.6a_S$. At the next order we recover the Lee-Yang form for the quantum depletion \cite{Lee-Yang,Fetter-Walecka} of the
molecular Bose gas with a coefficient that is only 6\% less than the correct asymptotic expression.

\textbf{(c)} At unitarity, both the zero-point motion of the collective modes and virtual
scattering of quasiparticles are important. Our numerical results (see Table I)
for the ground state energy, the gap and the speed of sound are in
good agreement with experimental data and Quantum Monte Carlo results; See Section VIII for details.

\textbf{(d)} The critical velocity $v_c$ across the crossover is maximum near unitarity, as previously predicted~\cite{rajdeep_vortex,Combescot} and as has been observed in experiments~\cite{vc_mit}.  We estimate an upper bound on $v_c$ using the Landau criterion, 
and find that quantum fluctuations considerably lower it with respect to mean field values; see Fig. 4.

\textbf{(e)} The results described above are obtained within a scheme in which the
Gaussian fluctuations do not feed back into the saddle point equation for the functional integral,
and they only contribute to the thermodynamic potential (\ref{thermodynamic_potential}).
This is a natural approximation within the functional integral framework, and we show in Appendix G
that it leads to exact answers in the simpler problem of the dilute repulsive Bose gas.
  
\textbf{(f)} To go beyond this approximation, we next include in Section IX the
self-consistent feedback of the Gaussian fluctuations into the gap equation. 
We find that this approach leads to several problems, some of which we can resolve. 
For instance, we show how the apparent violation of Goldstone's theorem in the self-consistent
scheme can be resolved  by going to an amplitude-phase representation of
the fluctuations. However, we point out that there are other problems which are not under control, 
such as the instability of the system in the extreme BEC limit.
Our detailed analysis of the theory with a 
gap equation modified by Gaussian corrections shows that imposing  
self-consistency does not necessarily lead to an improved approximation scheme.

In Section X we compare our approach and results with several other methods which have been used
to attack the same problem. Our approach has similarities with the $1/N$ expansion \cite{Sachdev,Leo}
but there are also differences which are discussed in Sec.~X. The equations solved in Section V are
the same as that obtained from the diagrammatic approach of Hu, Liu and Drummond \cite{Hu-Liu},
however our derivation is different and shows why it is natural not to renormalize the
saddle-point condition with fluctuation corrections. Further our approach also allows us to see
what the impact of going beyond this approximation is, as indicated in (f) above. This gives insights into 
problems faced in other self-consistent calculations \cite{Haussman}.
Our main conclusions are summarized at the end in Section XI.
Technical details of the calculations presented in the text are given in a series of six Appendices.
In a seventh Appendix we use illustrate the methods used in the text for paired Fermi superfluids
for the simpler case of a Bose superfluid. 

\section{II. Functional Integral Formalism }

We consider a system of fermions of two species, which we call ``spin'' $\sigma = \uparrow,\downarrow$, each of
mass $m$, described by the Hamiltonian density
\beq
H=\sib_{\sigma}(x)\left[-\frac{\nabla^2}{2m}-\mu\right]\psi_{\sigma}(x)-g\sib_{\up}(x)\sib_{\dn}(x)\psi_{\dn}(x)\psi_{\up}(x).
\label{hamiltonian}
\eeq
The first term has an implicit sum on the repeated index $\sigma$,
and the chemical potential $\mu$ is tuned to fix the average particle density $n = k_F^3/(3\pi^2)$
in a unit volume. Throughout the paper, we set $\hbar = k_B = 1$.

We consider the experimentally relevant case of a ``broad'' Feshbach resonance which can be adequately described within
a single-channel formulation of a dilute gas with $k_F r_{\rm eff} \ll 1$ where
$r_{\rm eff}$ is the range of the potential~\cite{wide}.  
The two-body interaction in (\ref{hamiltonian}) is described by a ``bare'' coupling constant $g$ and 
a momentum cut-off $\Lambda$, not explicitly shown above, which is of the order of $1/r_{\rm eff}$. 
The effective interaction at low-energies is completely described by
the s-wave scattering length $a_S$ for the two-body problem in vacuum. 
To obtain a given renormalized $a_S$, the bare coupling $g(\Lambda)$ must be tuned
using the relation
\beq
\frac{m}{4\pi a_S}=\frac{-1}{g(\Lambda)}+\sum_{|\kk| < \Lambda} \frac{1}{2\ep_\kk}
\label{scattering_length}
\eeq
where $\ep_\kk =  |\kk|^2/2m$. We will write
most of our equations in terms of the bare $g$, and only at the end 
we will use eq.(\ref{scattering_length}) to take the $\Lambda \to \infty$ limit 
and express the final results in terms of $a_S$.
(For a detailed discussion of justifying this regularization
procedure, we refer the reader to Sec.~IV of ref.~\cite{randeria90}).
 
The partition function $Z$ in the grand canonical ensemble at temperature $T$, chemical potential $\mu$
and in a unit volume, can be written as the imaginary time functional integral \cite{sademelo,engelbrecht97,dreschler}
over the Grassman fields $\sib$ and $\psi$ 
\beq
Z=\int D\sib_{\sigma} D\psi_{\sigma}\exp\left(-S_\psi\right)
\eeq
with the action
\beq
S_\psi=\int dx\ \left( \sib_{\sigma}(x)\partial_{\tau}\psi_{\sigma}(x)+H[\sib, \psi] \right).
\eeq
We use $x$ to denote $x=(\xx,\tau)$ where $\xx$ is the spatial coordinate and  
$\tau$ is imaginary time in the interval $0 \le \tau \le \beta$, where $\beta = 1/T$.
$\int dx = \int_0^{\beta}d \tau \int d^3\xx$ denotes an integral
over all space and over imaginary time.
Even though we are eventually interested in $T=0$, we find it convenient to use the finite $T$
Matsubara formalism and take $\beta \to \infty$ at the end. 

We next use a Hubbard Stratanovich transformation with an auxiliary field $\Delta(x)$ which couples to $\sib_{\up}(x)\sib_{\dn}(x)$
to obtain
\beq
Z=\int D\sib_\sigma D\psi_\sigma D\Delta D\Ds \exp\left(-S_{\psi,\Delta}\right).
\eeq
Using the spinor $\psi^{\dagger}(x)=\left(\sib_{\up}(x),\psi_{\dn}(x)\right)$ and 
its hermitian conjugate $\psi(x)$ the action can be written as
\beq
S_{\psi,\Delta}= \int dx \frac{|\Delta(x)|^2}{g}- \int dx dx' \psi^{\dagger}(x)\gr^{-1}(x,x')\psi(x')
\label{action}
\eeq
where the inverse Nambu-Gorkov Green's function $\gr^{-1}$ is given by
\beq
\left(
\begin{array}{clrr}%
        -\partial_{\tau}+\nabla^2/2m +\mu & {\ \ \ \ \ \ \ \Delta(x)} \\
        \Ds(x)  & -\partial_{\tau}-\nabla^2/2m -\mu
\end{array}
\right)\times\delta(x-x').
\label{nambuG}
\eeq
The functional integral is now quadratic in the fermion fields and these can be integrated out to
obtain 
\beq
Z=\int D\Delta D\Ds \exp\left(-S_\Delta\right)
\label{Z_Delta}
\eeq
with the action
\beq
S_\Delta =\int dx \frac{|\Delta(x)|^2}{g} - \int dx \ {\rm Tr}\ln \gr^{-1}[\Delta(x)]
\label{S_Delta}
\eeq
where the trace is over
two-dimensional Nambu space. Eq.~(\ref{Z_Delta}) is a formally exact expression for
$Z = \exp(-\beta\Omega)$, where $\Omega$ is the thermodynamic potential. 

\section{III. Mean Field Theory}

We briefly discuss the mean field theory \cite{leggett80,engelbrecht97} 
of the BCS-BEC crossover to introduce notation used throughout
the paper. Technical details highlighting aspects 
(such as convergence factors) which will be useful later
are given in Appendix A.

We begin by finding a spatially uniform, static saddle point $\Delta_0$
to the functional integral defined by eqs.~(\ref{Z_Delta},\ref{S_Delta}) 
This is determined by the gap equation
\beq
\delta S_\Delta / \delta\Delta_0 = 0,
\label{mf_saddle_point}
\eeq
where
\bqa
S_\Delta[\Delta_0]=\frac{\beta\Delta_0^2}{g} - \sum_{\kk,ik_n} {\rm Tr} \ln \gr_0^{-1}(k)\equiv S_0
\label{action_0}
\eqa
with
\beq
\gr^{-1}[\Delta_0] = 
\left(
\begin{array}{clrr}%
        ik_n-\xi_{\kk} & \ \ \ \ \Delta_0 \\
        \Delta_0  & ik_n+\xi_k
\end{array}
\right) \equiv \gr_0^{-1}(k).
\label{nambuG0}
\eeq
Here $ik_n = (2n+1)\pi/\beta$ are fermionic Matsubara frequencies,
and $\xi_\kk = \epsilon_\kk - \mu$ with $\epsilon_\kk = |\kk|^2/2m$.

After some straightforward algebra (see Appendix A)
the $T=0$ gap equation (\ref{mf_saddle_point}) can be finally written as
\beq
\frac{m}{4\pi a_S}=\sum_\kk\left[\frac{1}{2\ep_\kk}-\frac{1}{2E_\kk}\right]
\label{mf_gap_eq}
\eeq
where $E_\kk = \sqrt{\xi_\kk^2 + \Delta_0^2}$ and we have used 
eq.~(\ref{scattering_length}) to eliminate $g$ in favor of $a_S$.

To determine both $\Delta_0$ and the chemical potential $\mu$ we need to use
$n = - \partial\Omega / \partial\mu$, in addition to (\ref{mf_gap_eq}). 
At the level of the mean field (MF) approximation, the thermodynamic potential
is given by its saddle point estimate $\Omega_0 = S_0/\beta$ 
which leads to the $T=0$ MF number equation
\beq
n=\sum_\kk \left[1-\frac{\xi_\kk}{E_\kk}\right].
\label{mf_number_eq}
\eeq
Equations (\ref{mf_gap_eq}) and (\ref{mf_number_eq}) are the Leggett mean field
equations \cite{leggett80} for the $T=0$ BCS-BEC crossover which can be solved
to obtain the mean field values $\Delta_0$ and $\mu$ as a function of $(k_F a_S)^{-1}$
\cite{engelbrecht97}. 

Finally, we can obtain an explicit result for the MF thermodynamic potential at $T=0$ in terms of $\Delta_0$ and $\mu$.
We evaluate the Matsubara sum in eq.~(\ref{action_0}), take the $T=0$ limit, and use 
eq.~(\ref{scattering_length}) to obtain
\beq
\Omega_0 = -\frac{m}{4\pi a_S}\dl^2-\sum_\kk\left(E_\kk-\xi_\kk-\frac{1}{2}\frac{\dl^2}{\ep_\kk}\right).
\label{Omega_0} 
\eeq

\section{IV. Gaussian Fluctuations}

To go beyond the MF approximation
and include the effects of fluctuations, we write
\beq
\Delta(x) = \Delta_0 + \eta(x)
\eeq
where the complex bosonic field field $\eta(x)$ describes space-time
dependent fluctuations about the 
real, $(\xx,\tau)$-independent saddle point $\Delta_0$. 
We Fourier transform from $x= (\xx,\tau) \rightarrow q =(\qq,iq_l)$ 
where $iq_l = i 2\pi l/\beta$ is the Matsubara frequency for the bosonic $\eta$ fields. We then
write (\ref{nambuG}) as $\gr^{-1} = \gr_0^{-1} + {\bf K}$,
where $\gr_0^{-1}$ is defined in (\ref{nambuG0}) and
\beq
{\bf K}(k,k+q) = 
\left(
\begin{array}{clrr}%
        0 & \ \ \ \ \eta(q) \\
        \eta^*(-q)  & \ \ \ \ \ 0
\end{array}
\right)
\label{K-matrix}
\eeq
We next expand the action $S_\Delta$ to order $\eta^2$.
The first order term vanishes by the saddle point condition
(\ref{mf_saddle_point}) and we obtain
\bqa
S_{\Delta} = S_0 + S_g + \ldots
\label{gaussian_action}
\eqa
where the mean-field $S_0$ was defined in eq.~(\ref{action_0}).
The Gaussian piece has the form
\beq
S_g = {1 \over 2}\sum_{\qq,iq_l} \left( \eta^*(q),\eta(-q) \right)
\MM(q)
\left(
\begin{array}{clrr}%
        \eta(q) \\       
        \eta^*(-q)
\end{array}
\right).
\label{cartesian}
\eeq
The inverse fluctuation propagator $\MM$ is given by \cite{engelbrecht97,error}
\bqa
\no \MM_{11}(q) = \MM_{22}(-q) = \frac{1}{g}+\sum_{\kk,ik_n} \gr^0_{22}(k)\gr^0_{11}(k+q)\\
=\frac{1}{g}+\sum_\kk \left[\frac{u^2u'^2}{iq_l-E-E'} - \frac{v^2v'^2}{iq_l+E+E'}\right]
\label{m22}
\eqa
and
\bqa
\no \MM_{12}(q)= \MM_{21}(q) = \sum_{\kk,ik_n} \gr^0_{12}(k)\gr^0_{12}(k+q)\\
=\sum_\kk uvu'v'\left[\frac{1}{iq_l+E+E'}-\frac{1}{iq_l-E-E'}\right].
\label{m12}
\eqa
Here we use standard BCS notation
\beq
v_\kk^2 = 1 - u_\kk^2 = {1 \over 2}\left( 1 - \xi_\kk / E_\kk \right) 
\label{uk_vk}
\eeq
together with the abbreviations $u = u_\kk, v = v_\kk, E = E_\kk$ and
$u' = u_{\kk + \qq}, v' = v_{\kk+ \qq}, E' = E_{\kk + \qq}$.
The first line in (\ref{m22}) and (\ref{m12}) is valid at all temperatures, 
and the Matsubara sums lead to expressions involving Fermi functions $(1 - f - f')$ and $(f - f')$; 
see ref.\cite{engelbrecht97}. 
In the second line of (\ref{m22}) and (\ref{m12}) we only give results valid in the 
$T = 0$ limit where both $f(E)=f(E')=0$.
The factor of $1/g$ in (\ref{m22}) has to be regularized as usual using
eq.~(\ref{scattering_length}).
 
Integrating out the Gaussian fluctuations in
\bqa
Z \simeq \exp\left(-S_0\right) \int D\eta D\eta^\dagger \exp\left(- S_g\right) 
\label{ZactionM1}
\eqa
we obtain an improved estimate of the thermodynamic potential
\beq
\Omega \simeq \Omega_0 + \frac{1}{2\beta} \sum_{\qq,iq_l} \ln {\rm Det} \MM (q).
\label{Omega_1}
\eeq
where $\Omega_0$ was defined in (\ref{Omega_0}). 
{ We note that this result is true even for a \emph{non}-Hermitian matrix $\MM$,
provided its Hermitian part is positive definite~\cite{note on Gaussian integration}.  In our case, this condition corresponds to 
$M_{11} + M_{22} - 2 M_{12} >0$, which is true whenever (\ref{mf_gap_eq}) is satisfied. Physically this is related to an increase
in energy under a distortion of the phase, as can be seen from the analysis of Section IX.}

There is however a problem with this expression (\ref{Omega_1}), since it is actually ill-defined:
the Matsubara sum is divergent and 
we need appropriate convergence factors to make it meaningful
as discussed in detail in Appendix B.
We only write the final result here: 
\beq
\Omega \simeq \Omega_0 
+ \frac{1}{2\beta} \sum_{\qq,iq_l} \ln \left[{\MM_{11}(q) \over \MM_{22}(q)} {\rm Det} \MM (q)\right]e^{iq_l 0^+}.
\label{Omega_2}
\eeq

\begin{figure}[t!]
\includegraphics[width=3in]{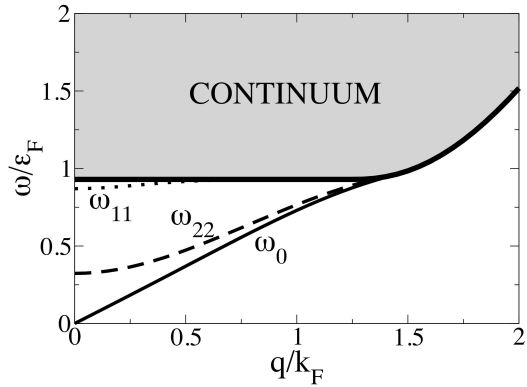}
\caption{Spectrum of excitations which contribute to the Gaussian correction to the thermodynamic potential (\ref{en_cont}).
The results shown correspond to unitarity $a_S = \infty$ with $\mu = 0.4 \,\epsilon_F$ and $\Delta_0 = 0.465\, \epsilon_F$. 
The full line shows the collective mode dispersion $\omega_0(q)$ (pole of $1/{\rm Det}\MM$) and the shaded region is the two-particle continuum
(branch cut). The dashed lines $\omega_{22}(q)$ and $\omega_{11}(q)$ are the zeros of , $M_{22}$, and $M_{11}$ respectively.}
\label{spectra0}
\end{figure}

In order to gain physical insight into what eq.~(\ref{Omega_2}) means, we will
analytically continue from Matsubara frequencies to real frequencies: $iq_l \to \omega + i0^+$.
Using standard manipulations (see Appendix B) the Gaussian 
part of the thermodynamic potential at $T=0$ can be written as 
$\Omega_g= - 1/2\sum_\qq\int_{-\infty}^0d\omega/\pi [\delta(\qq,\omega)-\delta_{22}(\qq,\omega)+\delta_{11}(\qq,\omega)]$.
Here $\delta$ is the phase of ${\rm Det} \MM$ defined by $\delta(\qq,\omega) = {\rm Im} \ln {\rm Det} \MM(\qq,\omega+i0^+)$
and $\delta_{22}$ and $\delta_{11}$ are the phases of $\MM_{22}$ and $\MM_{11}$ respectively.
The integral runs only over $\omega < 0$ because at $T=0$ the Bose factor $n_B(\omega) = - \Theta(-\omega)$.

The analytical structure of ${\rm Det} \MM(\qq,z)$ is as follows:
It has \emph{zeros} on the real axis at $z= \pm\omega_0(\qq)$, which correspond to 
\emph{poles} of the fluctuation propagator, and describe the spectrum of collective 
excitations. These excitations are oscillations of the phase of the order parameter as $q \rightarrow 0$, and is
the Goldstone mode arising from the broken symmetry 
in the superfluid state. We will show that $\omega_0(q) = c_s q$ as $q \to 0$ characteristic of a sound mode. 
In addition, at higher energies there are \emph{branch cuts} along the real axis at each $q$, with branch points at 
$\pm E_c(\qq)$ with $E_c(\qq) =\min(E_\kk+E_{\kk+\qq})$. These branch cuts represent 
the two-particle \emph{continuum} of states for scattering of gapped quasiparticles;
see the lower panel of Fig.~11 in Appendix D.

On the negative $\omega$ axis, $\MM_{22}(\qq,\omega)$ and $\MM_{11}(\qq,\omega)$ have 
zeros at $-\omega_{22}(\qq)$ and $-\omega_{11}(\qq)$ respectively, and each has
its own scattering continuum. Although the physical meaning of these quantities is less
clear, the role that they play in cutting off the ultraviolet divergences in eq.~(\ref{Omega_2}) will be 
clarified in detail below.

To illustrate these ideas, we show in Fig.~\ref{spectra0} 
the collective mode spectra and the two-particle continuum at the unitary point at which
$|a_S|$ diverges. Note that the collective mode frequency $\omega_0(\qq)$ is initially linear in q,
as expected, while the frequencies $\omega_{22}$ and $\omega_{11}$ have non-zero values in the limit $q\rightarrow 0$.  All of these frequencies eventually hit the two-particle continuum. Although we keep the integral in the Gaussian part
$\Omega_g$ over $\omega < 0$, as it appears in the algebra, we find it simpler to plot all spectra as
for \emph{positive} excitation energies in Fig.~\ref{spectra0} and subsequent figures.

We note in passing that even though the unitary Fermi gas is a very strongly interacting system,
nevertheless its collective mode spectrum does \emph{not} show a roton-like minimum observed in superfluid
Helium 4. We can understand this within a Feynman approach where the roton minimum arises from
a peak in the static structure factor characteristic of a \emph{liquid}, while here we are dealing
with a \emph{gas}, even if it is a very strongly interacting gas.

Next we explicitly separate out the collective mode and continuum contributions and write   
the thermodynamic potential $\Omega(T=0)= {\cal E} - \mu N$ as
\begin{eqnarray}
\Omega&=& \Omega_0 + \frac{1}{2}\sum_\qq\left[\omega_0(\qq)-\omega_{22}(\qq)+\omega_{11}(\qq)-E_c(\qq)\right]
\label{en_cont}\\
&-&\frac{1}{2\pi}\sum_\qq \displaystyle\int_{-\infty}^ {-E_c(\qq)} d\omega
\left[\delta(\qq,\omega)-\delta_{22}(\qq,\omega)+\delta_{11}(\qq,\omega)\right].
\nonumber
\end{eqnarray}
This is the full expression for the result (\ref{thermodynamic_potential}) in the Introduction.
 
The various contributions to the thermodynamic potential (\ref{en_cont}) are now much more transparent
compared with the Matsubara axis expression (\ref{Omega_2}).
$\Omega_0$ is the mean field contribution (\ref{Omega_0})
to the ground state energy ${\cal E}$. It may be interpreted as arising from filling up the 
negative energy $(-E_\kk)$ fermionic states of the BCS Hamiltonian, as is made clear in
Bogoliubov-deGennes theory.
The Gaussian contribution to ${\cal E}$ has three parts to it. The first part $\omega_0(\qq)/2$
comes from the zero-point motion of the collective mode. The second part, related to the
$\delta(\qq,\omega)$ terms, arises from virtual scattering of the fermionic quasiparticles
whose two-particle continuum begins at the energy $E_c(\qq)$.
The third set of contributions, related to the $\omega_{22}, \omega_{11}, \delta_{22}$ and $\delta_{11}$ terms,
come from the convergence factors of eq.~(\ref{Omega_2}) and are essential to get a 
finite answer for $\Omega$. 

To get a better feel for these various contributions it is useful to look at limiting cases.
In the BCS limit (Section VI) we will find that the quasiparticle scattering contribution gives the dominant contribution,
while in the BEC limit (Section VII) it is the zero point motion of the collective modes. The role of the
convergence factors is explained in more detail in Appendix B, and further insight will also be found
in the BEC limit.

\section{V. Results from Mean Field Theory plus Gaussian Fluctuations}

\begin{figure}[t!]
\includegraphics[width=3in]{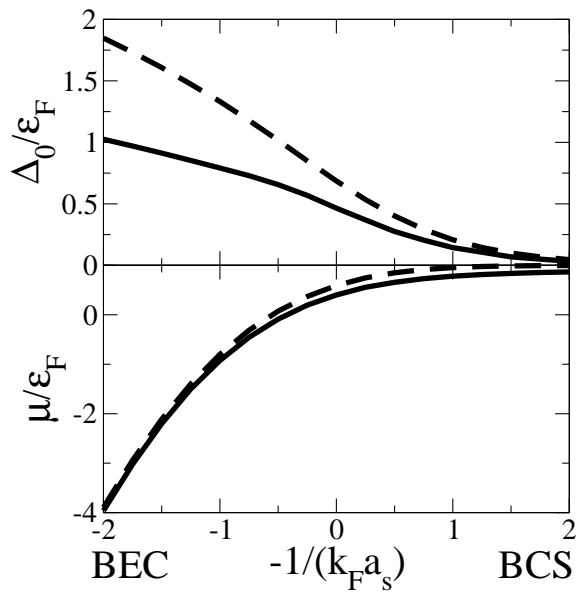}
\caption{Gap $\Delta_0$ and chemical potential $\mu$ as a function of $-(k_Fa_S)^{-1}$ across the BCS-BEC crossover.  
The dashed line is the mean field solution while the results of the calculation which includes Gaussian
fluctuations are shown as solid lines.}
\label{dlmu}
\end{figure}

Once the thermodynamic potential is obtained, we can find the chemical potential as well as all thermodynamical variables of the system.  
We must, however, first determine the uniform, static gap parameter $\Delta_0$. From eq.~(\ref{ZactionM1}), we see that the $\dl$ used in the expansion is the one that, for a given chemical potential, satisfies the {\em mean-field} saddle point equation (\ref{mf_saddle_point}), around which the action $S$ is expanded to quadratic order.
Thus, the gap and number equations 
\beq
\delta S_0 / \delta\Delta_0 = 0 \ \ \ {\rm and} \ \ \ n = - \partial\Omega / \partial\mu
\label{rpa_eqs}
\eeq
constitute the simplest theory which goes beyond the mean field approach and
is consistent with Goldstone's theorem (see below). As we shall see in this
and the next three Sections, this approach leads to very useful results and insights.
We note that even though the saddle point gap equation (\ref{mf_saddle_point}) used here
retains its mean-field \emph{form}, the \emph{values} of $\Delta_0$ and $\mu$ obtained from
the simultaneous solution of (\ref{rpa_eqs}) will deviate significantly from the mean field results 
(which are obtained using $\Omega_0$ of eq.~(\ref{Omega_0}) in the number equation).  
Moreover, as we show in Appendix G, an identical approach leads to
the known results in a different problem, that of a 
dilute repulsive Bose gas.
In Section IX, we will analyze a different scheme with a modified gap equation
which incorporates the self-consistent feedback of Gaussian fluctuations in the calculation of the saddle point, and show that it fails in some important aspects as an appropriate theory throughout the crossover.

In this Section we present the results on the following quantities 
across the BCS-BEC crossover obtained by adding Gaussian corrections 
to mean field theory: (i) the gap parameter $\Delta_0$ (ii) the chemical 
potential $\mu$ (iii) the ground state energy ${\cal E}$, (iv) the speed of 
sound $c_s$, and (v) the Landau critical velocity. In the next three Sections, we will discuss the asymptotic results
in the BCS and BEC limits and detailed numerical results at unitarity.

\begin{figure}[t]
\includegraphics[width=3in]{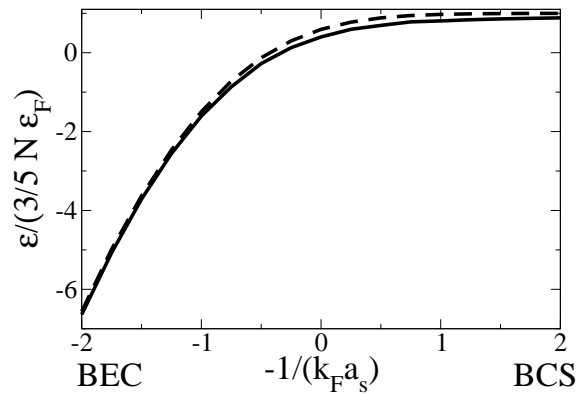}
\caption{Ground state energy per particle ${\cal E}$ in units of the non-interacting 
result $3\ep_f/5$ as a function of $-1/(k_Fa_S)$. The difference
between the mean field result (dashed line) and the Gaussian fluctuation calculation (solid line)
is small and more clearly shown in Fig.~\ref{BCS_regime.fig} (BCS limit), Fig.~\ref{BEC_regime.fig} (BEC limit)
and Table I (unitarity).
}
\label{gs_en_fig}
\end{figure}
In order to obtain $\Delta_0$ and $\mu$ from (\ref{rpa_eqs}), we solve the gap equation 
(\ref{mf_gap_eq}) for $\dl(\mu)$ together with the number equation written as
\beq
n=-\frac{\partial \Omega_0}{\partial\mu}-\frac{\partial \Omega_g[\mu,\dl(\mu)]}{\partial\mu}.
\label{new_number-eq}
\eeq 
Note that the thermodynamic $\mu$-derivative 
(keeping volume and $T=0$ fixed) in (\ref{new_number-eq}) must take into account the
$\mu$-dependence of the saddle point $\dl(\mu)$.
(An analogous point for the possibly more familiar case of the dilute Bose gas is emphasized
in Appendix G). 

To solve the above number equations we must numerically evaluate $\Omega_g[\mu,\dl(\mu)]$.
Even though the real-frequency representation (\ref{en_cont}) gives 
physical insight, we find it simpler to numerically evaluate $\Omega_g$ on the Matsubara axis, 
as described in Appendix C. Finally, we calculate $F(\mu)= \Omega_0 + \Omega_g[\mu, \Delta_0(\mu)] + \mu n$ 
and look for an extremum (maximum) as a function of the chemical potential $\mu$.

The gap $\dl$ and the chemical potential $\mu$ are plotted as a function of 
$-(k_Fa_S)^{-1}$ in Fig.~\ref{dlmu}, where the dashed line is the MF value
for comparison. As expected, the inclusion of fluctuations reduces the value of $\Delta_0$.
We note that the auxiliary field $\Delta_0$ continues to determine the
energy gap $E_g = \Delta_0$ for $\mu > 0$ and $E_g = \sqrt{|\mu|^2 + \Delta_0^2}$ for $\mu < 0$
(just as in MF theory), so long as we ignore the feedback of the fluctuations on the single-particle propagator.

The ground state energy of the system is obtained from the thermodynamical potential using 
${\cal E} = \Omega(T=0) + \mu n$ is plotted in Fig.~\ref{gs_en_fig}. We see that although the difference
between the MF and Gaussian results is quite small, fluctuations reduce the ground state energy
through the entire crossover. The quantitative superiority and the physical insights of the Gaussian
results are discussed in detail later: see Fig.~\ref{BCS_regime.fig} for the BCS limit, Fig.~\ref{BEC_regime.fig} for the BEC limit
and Table I for the results at unitarity.

We next compute the speed of sound through the BCS-BEC crossover.
First, we emphasize that Goldstone's theorem is necessarily obeyed by the theory
defined by (\ref{rpa_eqs}); this is in contrast to the self-consistent calculation
to be described in Section IX. The existence of a 
zero energy Goldstone mode is guaranteed by the
form of the gap equation (\ref{mf_saddle_point}) which implies that
${\rm Det}M(\qq=0,\omega=0) = 0$. To see this fact, note that
we can write
\beq
{\rm Det}\MM(0,0)=\left[\frac{1}{g}+\sum_k {\rm Det}\gr\right]\left[\frac{1}{g}+\sum_k\gr_{22}\gr_{11}+\gr_{12}^2\right] 
\label{detM_00}  
\eeq
and the saddle-point condition is $1/g=-\sum_k {\rm Det} \gr$. 

\begin{figure}[t]
\includegraphics[width=3in]{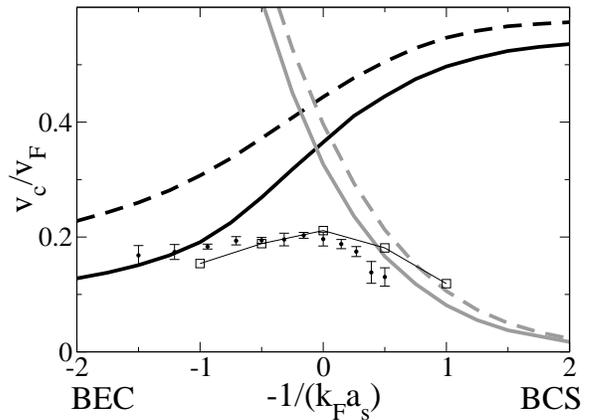}
\caption{The Landau critical velocity as a function of $-1/(k_F a_S)$ in the BCS-BEC crossover is given
by ${\rm min}\{{\cal E}(k)/k\}$. The black lines represent the speed of sound $c_s$, and
the gray lines the pair breaking estimate obtained using (\ref{vc}). In each case 
the dashed line is the result using the MF gap and chemical potential, while 
the solid line is the result obtained after inclusion of Gaussian fluctuations. { The 
black data points (with error bars) are the experimental results from ref.~\cite{vc_mit}. The open squares
are the result of a Bogoliubov-deGennes vortex calculation from ref.~\cite{rajdeep_vortex}.}
}
\label{vc_fig}
\end{figure}

The collective mode spectrum has the form $\omega_0(\qq)  = c_s |\qq |$ for $q\rightarrow 0$ where
$c_s$ is the speed of sound.  We calculate $c_s$ following the approach of 
\cite{engelbrecht97}; we include the expressions here to correct a typographical error \cite{error} 
in that reference.  
Rotating the frequency from the Matsubara axis to the real line ($iq_l \rightarrow -\omega$) and expanding to quadratic order in both momentum and frequency, we get $M_{11}(-\omega, \qq) = (A + 2 B\omega + (C +Q) |\qq|^2 - (D+R) \omega^2)/2$, $M_{22}(-\omega, \qq) = M_{11}(\omega, {\bf q})$ and 
$M_{12}(-\omega, \qq) = (A + (C - Q) |\qq|^2 - (D-R) \omega^2)/2$.  Here $A = \sum_\kk \Delta_0^2/2E^3$, $B = \sum_\kk \xi/2E^3$, $C = \sum_\kk \{ (1-3X)\xi/m - [1-10X(1-X)]Y\}/8E^3$, $D=\sum_\kk (1-X)/8E^3$, $Q = \sum_\kk \{\xi/m - (1-3X)Y\}/8E^3$, and finally $R = \sum_\kk 1/8E^3$, with the notation $X= \Delta_0^2/E^2$, and $Y = |\kk|^2/3m^2$.
We thus obtain
\begin{equation}
\label{speed of sound}
c_s = \sqrt{Q /\left[ B^2/A + R \right]}.
\end{equation}
The results for the speed of sound across the BCS-BEC crossover are shown as the black curve
in Fig.~\ref{vc_fig}.  
The solid line is the result
obtained after inclusion of Gaussian fluctuations, while the dashed line is the result using the MF
values for $\Delta_0$ and $\mu$. The other curves shown in this Figure are discussed below.

To conclude this Section, we turn to the calculation of the Landau critical velocity $v_c$ as a function
of $-1/(k_F a_S)$. As we have seen, most observables -- gap, chemical potential, ground state energy,
speed of sound -- are monotonic functions of $1/k_F a_S$ through the crossover. The same is true of
the transition temperature $T_c$ which shows a slight maximum near unitarity \cite{sademelo} but is essentially
the same, of order $0.2 \epsilon_f$ for all positive scattering lengths, i.e., to the BEC-side of unitarity.
In other words there seems to be nothing particularly dramatic about the properties of the most strongly interacting unitary 
regime. However, as first pointed out in ref.~\cite{rajdeep_vortex} based on the study of the current flow
around a vortex, the critical velocity as a function of $1/k_F a_S$ has a strongly nonmonotic behavior
through the crossover with a pronounced peak at (or close to) unitarity. The reason for
this behavior is that very different excitations are responsible for the destruction of superfluidity \cite{rajdeep_vortex,Combescot}:
breaking of pairs on the BCS side and generation of phonons on the BEC side of unitarity.  

To understand this better and to compare with recent experimental data, we use the
Landau criterion which gives an upper bound on the critical velocity of the form
$v_c = {\rm min}\{{\cal E}(k)/k\}$ where ${\cal E}(k)$ is the energy of an excitation carrying momentum $k$.
We separately consider single-particle (fermionic) and collective (bosonic) excitations. 
For single-particle (s.p.) excitations, the excitation energy is $E_\kk = \sqrt{(\epsilon_\kk - \mu)^2 + \dl^2}$
which then leads to the $v_c$ estimate
\beq
\left({v_c \over v_f}\right)_{\rm s.p.} 
= \left[{{\left(\sqrt{\mu^2 + \dl^2} - \mu\right)^2 + \dl^2}\over{4\epsilon_f\sqrt{\mu^2 + \dl^2}}}\right]^{1/2}.
\label{vc}
\eeq
As we shall see, this is most relevant on the BCS side of unitarity, and in the BCS limit $\mu \gg \dl$, it simply
reduces to the well-known result for pair-breaking $v_c \simeq \dl/k_F$. The pair breaking estimate
of eq.~(\ref{vc}) is plotted in gray in Fig.~\ref{vc_fig}. The dashed gray line is the result using the MF gap and 
chemical potential, while the solid gray line is the result obtained after inclusion of Gaussian fluctuations. 

For collective excitations, we find that the Landau critical velocity is given by the slope
of the tangent to the $\omega_0(k)$ curve. Since there is no roton dip for the superfluid Fermi gas
(as already remarked), one simply gets the speed of sound:
\beq
\left(v_c \right)_{\rm coll.} = c_s
\label{vc_cs}
\eeq

The actual critical velocity is then bounded above  by the minimum of 
the single particle $\left(v_c \right)_{\rm s.p.}$ and collective $\left(v_c \right)_{\rm coll.}$.An estimate of the Landau critical velocity at the mean field level, which corresponds to the dashed curves in Fig. 4, was given in \cite{Combescot}.  We find that quantum fluctuations lead an appreciable reduction in $v_c$, as seen in the full curves in the figure.  We also plot the results of a recent experimental study
of the critical velocity \cite{vc_mit}, for which $v_f = 30$ mm/s.  The theoretical predictions of the
nonmonotocity of $v_c$ with a peak around unitarity and their experimental confirmation 
show that the unitary Fermi gas is the most robust superfluid in the entire crossover.

\section{VI. BCS Limit: Hartree shift and Fermi liquid corrections}
We now describe in detail the BCS limit solution for $1/k_F a_S \to - \infty$. 
We will show that the collective mode contribution
in eq.~(\ref{en_cont}) is very small because of phase space restrictions, and the dominant correction
to $\Omega(T=0)$ comes from virtual scattering of quasiparticles. We find that the Gaussian theory
recovers the well-known ``normal state'' correction to the ground state energy of a dilute Fermi gas
originally studied by Huang, Yang and Lee and by Galitskii. The leading term in this correction, which is 
of order $k_F a_S$, is the Hartree shift of the ground state energy. We note that the BCS mean field ground
state energy differs from the free Fermi gas by a condensation energy of order $\Delta_0^2/\epsilon_f$ and represents an
exponentially small correction of order $\exp(- 1/k_F|a_S|)$ relative to $\epsilon_f$. In contrast,
the Gaussian fluctuation contributions will be found to be power-law corrections in $k_F |a_S|$.  

\begin{figure}[t!]
\includegraphics[height=2in]{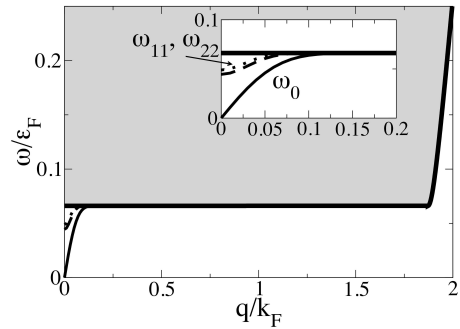}
\caption{Spectrum of excitations which contribute to the Gaussian correction to the thermodynamic potential (\ref{en_cont}).
The results shown correspond to the BCS regime $1/(k_F a_S) = -2$, with $\mu = 0.867 \,\epsilon_f$ and $\Delta_0 = 0.0311\, \epsilon_f$. 
The full line shows the collective mode dispersion $\omega_0(q)$ (pole of $1/{\rm Det}\MM$) and the shaded region is the two-particle continuum
(branch cut). The dashed lines $\omega_{22}(q)$ and $\omega_{11}(q)$ are the zeros of , $M_{22}$, and $M_{11}$ respectively,
which coincide in the particle-hole symmetric BCS limit. The inset shows long-wavelength, low-energy spectra.}
\label{spectra_2}
\end{figure}

The BCS limit 
is characterized by an exponentially small gap $\dl$ and $\mu \approx \epsilon_f$.  
The spectrum of collective excitations found in the BCS regime is shown in Fig. \ref{spectra_2} 
for $1/(k_F a_S) = -2$ with $\mu = 0.867\, \epsilon_f$ and $\dl = 0.0311 \, \epsilon_f$.  
In this case, at $q=0$ the continuum starts at a frequency equal to $2\dl$, and the collective modes are restricted to a 
small frequency interval, magnified in the inset. Due to particle-hole symmetry in the BCS limit, 
the zeros $\omega_{22}$ and $\omega_{11}$ coincide and hence do not contribute to the energy (\ref{en_cont}).  The speed of sound in this limit becomes 
$c_s\simeq v_f/\sqrt{3}$~\cite{Anderson} as we show at the end of this section.

The MF ground state energy of the superfluid state is given
by the well known BCS result ${\cal E}_0 = 3n\epsilon_f/5 - (3n\Delta_0^2/8\epsilon_f)$. 
We first show that the contribution of the zero-point
motion of the collective modes to $\Omega$ is exponentially smaller than the 
(already small) MF condensation energy, and may be neglected.
The momentum $q_c$ where the pole hits the continuum at $2 \Delta_0$ is given by 
$q_c\sim \dl/v_f\sim \xi^{-1}$ where $\xi$ is the correlation length.
Thus the phase space available for the collective mode contribution  is tiny
because $q_c \ll k_F$. 
The contribution per particle to (\ref{en_cont}) coming from the poles
is seen to be $\sim \ep_f (k_F\xi)^{-4}\sim\ep_f(\dl/\ep_f)^4$ which
is negligible.

We next turn to the continuum contribution to (\ref{en_cont}). In the BCS limit it is 
justified to set $\dl=0$ here to get the leading order terms in the ground state energy. Any 
corrections due to non-zero $\dl$ are at least down by a factor of ${\cal O}(\dl^2/\ep_f)$ and 
thus negligible. As already remarked, the ground state is a superfluid which 
leads to a MF energy reduction of ${\cal O}(\dl^2/\ep_f)$, with respect 
to the energy that we will calculate. This exponentially small 
contribution is vital to get a stable ground state, but once this is 
done, we can set the gap to zero in computing leading order corrections to the ground state energy,
as explained in Appendix D.

In the BCS limit, our result for $\Omega_g$ is exactly of the form of the well-known results for the 
dilute Fermi gas in its \emph{normal} state~\cite{Fetter-Walecka}, but with $a_S < 0$,
as shown in Appendix D. We find that 
the total energy is given by 
\begin{eqnarray}\label{energy_BCS.formula}
{{\cal E} \over {n \epsilon_f}} &=& {3 \over 5} - {3\dl^2 \over 8\ep_f^2} \\
& &+ {2 \over 3\pi} k_Fa_S
+ {4(11 - 2\ln 2)\over 35\pi^2} (k_Fa_S)^2\nonumber
+ \dots
\end{eqnarray}
and the chemical potential 
\begin{eqnarray}\label{mu_BCS.formula}
{\mu \over \epsilon_f} &=&1 - {\cal O}(\dl/\ep_f)^2\\
& &+ {4 \over 3\pi} k_Fa_S
+ {4(11 - 2\ln 2)\over 15\pi^2} (k_Fa_S)^2 \nonumber
+ \dots
\end{eqnarray} 
The first order term is the Hartree term while the second order contributions have the same
form as those obtained by Huang, Yang and Lee and by Galitskii ~\cite{galitskii,Lee-Yang} for the 
dilute Fermi gas, except that in our case $a_S < 0$.

It is worth commenting that, although it is customary to think of the Hartree term as a ``mean field shift'',
it arises in our approach as the first term in the \emph{fluctuation} correction to the saddle point thermodynamic potential
in the BCS limit. This is also seen clearly from Fig.~\ref{lyg} in Appendix D where the first
diagram is clearly the Hartree term. Once we move away from the BCS limit toward
unitarity, the gaussian fluctuation contribution is still well defined even in absence of a small parameter, 
however a ``Hartree term'' becomes hard to identify.

\begin{figure}[t] 
\includegraphics[height=3.2in]{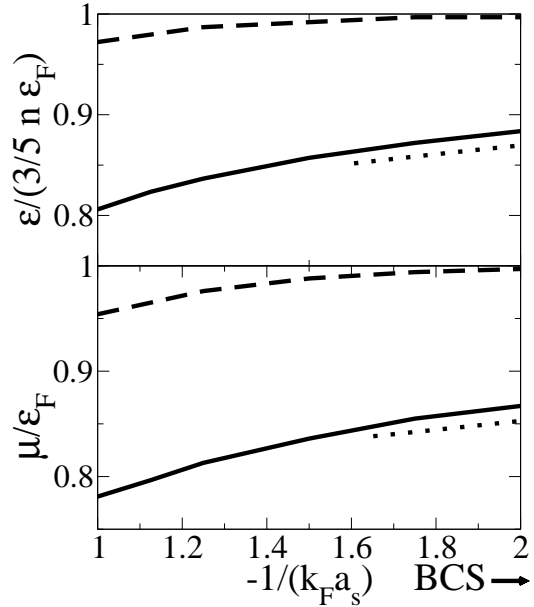}
\caption{Energy and chemical potential in the BCS regime, for $-(k_Fa_S)^{-1}$ between 1 and 2.  The solid lines are our calculations
including Gaussian fluctuations, the dashed lines the mean field values, and the dotted lines the results of formulae (\ref{energy_BCS.formula}) and (\ref{mu_BCS.formula}), in the top and bottom panels, respectively.}\label{BCS_regime.fig}
\end{figure}

In Fig.~\ref{BCS_regime.fig} we plot the energy per particle as well as the chemical potential as function
of interaction in the BCS regime: $-2 < (k_Fa_S)^{-1} < -1$.  We have also included the mean field result and 
the asymptotic values given by (\ref{energy_BCS.formula}) and (\ref{mu_BCS.formula}).  As we can see, quantum fluctuations 
reduce the energy and the chemical potential, in a way that is consistent with the corrections obtained from our analysis.

In order to calculate the speed of sound, one needs to look at the 
slope of the pole dispersions, which in a superfluid is given 
by $c=\sqrt{\rho_s/\kappa}$, where $\rho_s$ is the superfluid density 
and $\kappa=m \partial n/\partial\mu$ is the compressibility of the system.
In a homogeneous system, Galilean invariance implies that $\rho_s=n$ at $T=0$. 
Keeping only the Hartree term in the formula for the chemical potential, 
we obtain to linear order in the scattering length $c_s=v_f/\sqrt{3} (1+ {1\over \pi} k_Fa_S)$,
which was first obtained for BCS superconductors by Anderson~\cite{Anderson}.

We conclude this Section with a comment on the Gorkov Melik-Barkhudarov 
correction \cite{gmb}
which enters the pre-exponential factor in the BCS expression for the gap. It arises from the renormalization 
of the effective attraction by particle-hole excitations in the medium. Such effects are \emph{not} accounted
for in our theory which effectively considers only particle-particle channel diagrams. At the present time we do not know
of any theory which recovers this correction in the extreme BCS limit and shows how this correction evolves
through unitarity. 

\section{VII. BEC Limit: Dimer scattering and Lee-Yang corrections}

We next describe in detail the BEC limit solution for $1/k_F a_S \to + \infty$. We will see that the zero-point
motion of collective modes entirely dominates over the continuum contribution in the thermodynamic potential. 
We will find that this leads to two important effects in the BEC limit: first, a reduction of the effective
dimer-dimer scattering length relative to its mean field value of $2a_S$ and second, a Lee-Yang correction to
the equation of state of the dilute gas of dimers. While our theory is able to obtain both these effects semi-quantitatively, 
it does not give the exact asymptotic answers. The scattering length for bosonic molecules, or dimers, is found to be $\simeq 0.55 a_S$,
while the exact solution of the four-body problem yields $0.6 a_S$~\cite{petrov}, and the coefficient of the Lee-Yang correction 
is only 6\% smaller than the exact result.

\begin{figure}[t!]
\includegraphics[width=3in]{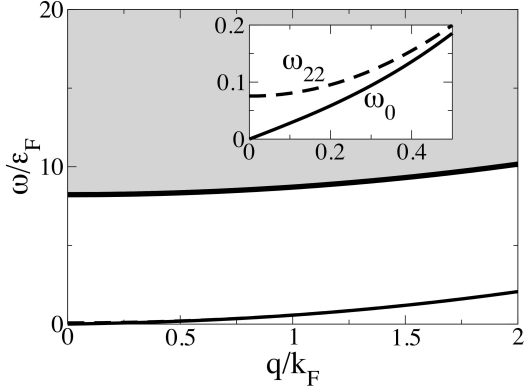}
\caption{Spectrum of excitations which contribute to the Gaussian correction to the thermodynamic potential (\ref{en_cont}).
The results shown correspond to the BEC regime with $1/(k_F a_S) = 2$, with $\mu = -3.967 \epsilon_F$ and $\Delta_0 = 1.025 \epsilon_F$..
The full line shows the collective mode dispersion $\omega_0(q)$ (pole of $1/{\rm Det}\MM$) and the shaded region is the two-particle continuum
(branch cut). The dashed line $\omega_{22}(q)$ are the zeros of $M_{22}$. We do not show
$\omega_{11}(q)$, the zeros of $M_{11}$, because they coincide with the continuum in the BEC limit.
The inset shows the long wavelength, low energy spectra.}
\label{spectra2.fig}
\end{figure}

In the BEC limit the chemical potential is large and negative and, to leading order, goes to one half of the binding energy of the 
molecules: $\mu = - E_b/2$ where $E_b=1/ma_S^2$. The spectrum for collective excitations is shown in Fig.~\ref{spectra2.fig} 
for $1/k_Fa_S = 2$.  The two-particle continuum then sits at a very high energy; at $|\qq|=0$ it begins at an energy of
$2\sqrt{|\mu|^2 + \dl^2} \approx E_b$. 
The virtual scattering of the very high energy fermionic excitations makes
a negligible contribution to the thermodynamic potential in the molecular BEC limit.

The Gaussian contribution is then entirely 
dominated by the low frequency collective mode $\omega_0(q)$, which is the Bogoliubov excitation
of the molecular Bose gas, and the mode $\omega_{22}(q)$ coming from the convergence factor. These
modes are shown in more detail in the inset of Fig.~\ref{spectra2.fig}.
Note that for $\mu < 0$, $\omega_{11}$, the zeros of $M_{11}$, are pushed to the continuum
and do not enter the calculation.
The thermodynamic potential then simplifies to 
$\Omega_g \simeq 1/2\sum_\qq[\omega_0(\qq)-\omega_{22}(\qq)]$. In the BEC limit the 
pole always remains below the continuum for all $q$ 
and the phase space for the zero-point oscillations extend formally to 
$q = \infty$. 

This raises the question: how can we get a finite answer for the sum in $\Omega_g$?
The dispersion of the pole is given by
the standard Bogoliubov expression
$\omega_0(\qq)=\sqrt{c_s^2\qq^2+(\qq^2/2m_b)^2}$, where 
$m_b=2m$ is the mass of the bosonic molecule as shown in
ref.~\cite{engelbrecht97}. The large $\qq$ limit 
of this dispersion is just the kinetic energy of the bosonic molecule, which 
should \emph{not} be part of the zero-point motion of the fluctuations. This is
exactly where the convergence factors come in. One finds that 
$\omega_{22}(\qq) \rightarrow \qq^2/2m_b$ for large $\qq$ and cancels
the contribution of the free boson dispersion. Thus the convergence 
factor gives rise to manifestly finite results by eliminating the 
free particle dispersion of the pole spectrum from contributing to the 
zero-point motion of the phase fluctuations. (See Appendix G for
an analogous discussion for the dilute Bose gas.)

\begin{figure}[t] 
\includegraphics[width=3in]{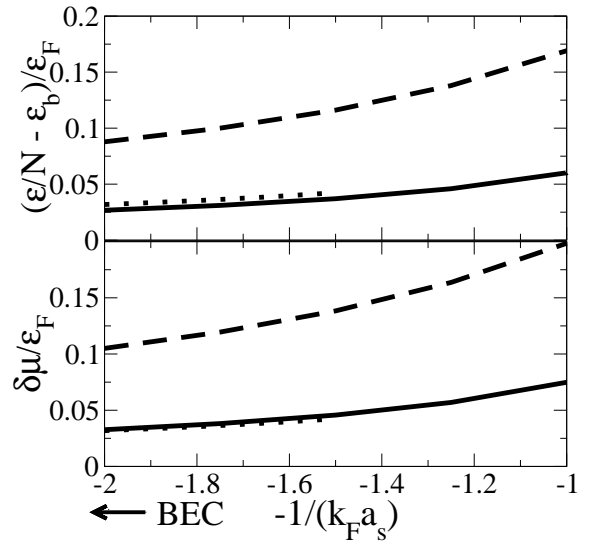}
\caption{
Energy and chemical potential in the BEC regime, for $(k_Fa_S)^{-1}$ between 1 and 2.  The solid lines are our calculations
including Gaussian fluctuations, the dashed lines the mean field result, and the dotted lines the leading order result for
a dilute Bose gas with an effective repulsion $a_b = 0.6 a_S$.}\label{BEC_regime.fig}
\end{figure}

We now calculate the leading order correction to the mean field
thermodynamic potential in the BEC limit, which will in turn determine the 
effective interaction between the bosonic molecules. The idea is to use an
expansion in the small parameter $\Delta_0/|\mu| \ll 1$ in the BEC limit.
The calculation is most easily done on the Matsubara axis as detailed in Appendix E. 
Here we only quote the leading order result:
\beq
\Omega_g \simeq  -{\alpha\over 256 \pi} (2m)^{3/2}\frac{\dl^4}{|\mu|^{3/2}}
\label{Omega_bec}
\eeq 
where we have included the factor of $256 \pi$ in order to simplify later expressions.  
The $m, \Delta_0$ and $\mu$ dependence can be determined analytically and the
dimensionless prefactor $\alpha =2.61$ has to be evaluated by a
numerical integration, as shown in Appendix E.

To find the effective scattering length between the molecules we  
calculate the shift in the bosonic chemical potential as a result of
interactions. Toward this end we proceed as follows. 
We expand the gap equation (\ref{mf_gap_eq})
in powers of $\Delta_0/|\mu| \ll 1$ to recover the 
Gross-Pitaevskii equation for bosons. We then get
\beq
\frac{1}{a_S}=\sqrt{2m|\mu|}\left(1+\frac{1}{16}\frac{\dl^2}{|\mu|^2}\right).
\label{gp1}
\eeq
As noted earlier, the leading order result is $\mu=-1/(2ma_S^2)$. The above result
allows us to relate the next order correction in $\mu$ to $\Delta_0$.
We find
\beq
\mu=-1/(2ma_S^2)+\delta\mu, \ \ \ \  \delta\mu = ma_S^2\dl^2/4.
\label{delta_mu}
\eeq

Next we determine $\Delta_0$ from the number equation (\ref{new_number-eq}).
We find:
\beq
n=n_{\rm MF}-\left(\frac{\partial \Omega_g}{\partial \mu}\right)_{\dl} -
\left(\frac{\partial \Omega_g}{\partial \dl^2}\right)_\mu\frac{\partial \dl^2}{\partial \mu}
\label{n-bec}
\eeq 
where $n_{\rm MF}=\dl^2m^{3/2}/(4\sqrt{2}\pi|\mu|^{1/2})$ is the MF result in the BEC limit.
(See Appendix G for an analogous discussion of keeping track of 
the $\mu$ dependence of the saddle point for a dilute Bose gas.)
Using (\ref{Omega_bec}) we get 
$\partial \Omega_g/\partial \mu=(\alpha/256\pi)(2m)^{3/2}(3/2)\dl^4/|\mu|^{5/2}$ 
and 
{ the last term in (\ref{n-bec}) is given by}
$(\alpha/32\pi)(2m)^{3/2}\dl^2/(|\mu|^{3/2}ma_S^2)$. It is easily seen that 
$-\partial \Omega_g/\partial \mu\sim (k_Fa_S)^3$ and can be 
neglected in the BEC limit. Now, using $|\mu|=1/(2ma_S^2)$ to leading order, 
we get
\beq
\dl^2 = \left(\frac{16}{3\pi}\right)\ep_f^2\frac{1}{k_Fa_S}\left(\frac{1}{1+\alpha}\right)
\eeq
This leads to $\delta\mu=(2/3\pi)\ep_fk_Fa_S(1/(1+\alpha))$. Comparing 
the chemical potential for the bosons $\mu_b=2\delta\mu$ with the weakly interacting Bose gas
result $\mu_b=4\pi a_bn_b/m_b$ with 
$n_b=n/2$ and $m_b=2m$, we get the effective scattering length for the bosons 
to be 
\beq
a_b= 2a_S/(1+\alpha) \simeq 0.55a_S,
\eeq
using our numerical result $\alpha = 2.61$. This result for $a_b$ is identical with 
the one obtained by Hu {\it et al.} \cite{Hu-Liu}; see Section X for further discussion.

Going beyond the leading order term we find that the next order 
correction to the chemical potential is of order $(n_b a_b^3)^{1/2}(n_b a_b/m_b)$,
which has the same form as the Lee-Yang corrections for a 
weakly repulsive Bose gas~\cite{Lee-Yang}. To see this
we analyze our numerical results for $\Omega$ in the BEC limit as follows. 
Scaling out energies with $1/(2ma_S^2)$ and lengths with $a_S$, we
find that we can fit $\Omega_g$ to the functional form
$\Omega_g = (1/2ma_S^2)a_S^{-3}[A(\widetilde{\delta\mu})^2 + B(\widetilde{\delta\mu})^{5/2} + \ldots]$
where $\widetilde{\delta\mu} = 2ma_S^2 \mu - 1$.  {Solving for the molecular chemical potential $\mu_b = 2 \delta \mu$ we obtain
\begin{equation}
\mu_b = {4\pi n_b a_b\over m_b} \left [ 1+\gamma \, {32\over 3\sqrt{\pi}}\, (n_ba_b^3)^{1/2} + \cdots \right ]
\end{equation}
where we find that the coefficient $\gamma = 0.94$ is 6\% smaller than the Lee-Yang result $\gamma=1$ (see Appendix G).}
%

In  Fig.~\ref{BEC_regime.fig} we show the total energy and the chemical potential for interactions in the BEC
regime $1 < 1/k_Fa_S < 2$.  For comparison we also show the mean field results (dashed lines) and the 
leading order result for a gas of bosons interacting with $a_b = 0.6 a_S$ (dotted lines).

\section{VIII. Unitarity}

At unitarity ($1/k_Fa_S=0$), there is no small parameter
and the problem can only be solved numerically. At this point both
the pole and the continuum corrections 
are of comparable magnitude in the thermodynamic potential. We present the 
results for the gap, chemical potential, ground state energy and the speed of sound and compare 
with quantum Monte Carlo and experimental results in Table~\ref{table_unit}.

At unitarity $|a_S|$ diverges and this leads to the 
concept of universality, i.e., all energies scale with the Fermi energy 
$\ep_f$ and all lengths scale with $k_F^{-1}$. A consequence of universality 
is the relation between the ground state energy per particle and the 
chemical potential $\mu=(5/3)({\cal E}/N)$, which acts as a check on our 
numerical calculation. The ground state energy is generally written in terms 
of the non-interacting energy ${\cal E}/N=(1+\beta)  3\ep_f/5$. 
(Note that in this Section $\beta$ is used to denote the universal interaction
correction to the ground state energy, and {\it not} the inverse temperature).
We obtain a numerical value of $\beta = -0.598(1)$. The mean field theory gives $\beta=-0.41$~\cite{engelbrecht97}, while 
quantum Monte Carlo methods give a $\beta$ of $-0.56$~\cite{Carlson}. The experimentally 
obtained values range from $-0.68$ to $-0.49$~\cite{betaGrimm,betaENS,betaThomas}. 
We thus see that at unitarity the Gaussian quantum fluctuations -- zero-point motion of collective modes
and virtual quasiparticle scattering -- account for most of the difference between the 
exact ground state energy (i.e., that obtained from QMC or experiments) and the simple mean field estimate.

The speed of sound is obtained from the dispersion of the pole 
$\omega=c_s|\qq|$ at small momenta, or alternatively in our theory from eq.~(\ref{speed of sound}).  
However, we can also calculate the speed of sound once we know the equation of state.  
Using that $\partial \mu / \partial n = (2\mu)/(3n)$, we arrive at the expression 
$c_s/v_f = \sqrt{(1+\beta) /3}$.  Our theory predicts a speed of sound at unitarity of 
$c_s=0.37 \, v_f$, using either one of the mentioned methods. For comparison, the answer \cite{engelbrecht97}
obtained by using the mean field gap and chemical potential is $0.44 v_f$. 
The Quantum Monte Carlo estimate is $c_s = 0.38 v_f$ while the experimentally measured value of the 
speed of sound at unitarity is  $0.38 v_f$~\cite{Thomas sound}. 

We also show in the last column of Table I various gap estimates.
The inclusion of fluctuations reduces the value of $\Delta_0$ relative to
the MF estimate. As already noted above, $\Delta_0$ continues to determine the
energy gap $E_g = \Delta_0$ for $\mu > 0$, so long as we ignore the feedback of the fluctuations on 
the single-particle propagator. We do not include an experimental value for the
energy gap as we believe that it is not clear how to quantitatively extract this from
rf spectroscopy data, taking into account the interactions between atoms in the final and
initial states~\cite{note on rf spectroscopy}. 

\section{IX. Self-consistent Feedback of Gaussian Fluctuations on the Saddle-point}

In this Section we describe how one can include the feedback of the Gaussian fluctuations 
on the saddle-point equation in a self-consistent manner. We show here that it is straightforward
to accomplish this using a fluctuation formalism similar to the one used in Section IV. 
However, one finds, quite generally, that Goldstone's theorem is violated
if one uses such a Cartesian representation for the fluctuations, as soon as one modifies the
saddle point equation. We next show that a polar representation of the fluctuations in terms of the
amplitude and phase of the auxiliary field allows one to recover the Goldstone mode, even when
the saddle point condition is modified away from its mean field form. There is still a problem
with obtaining an ultraviolet convergent expression for the thermodynamic potential in terms of
the fluctuations. The polar representation respects Goldstone's theorem at low energies but has unacceptable
high energy properties, while the Cartesian representation violates Goldstone in the infrared but is well 
controlled in the ultraviolet. We thus construct a hybrid representation which interpolates between the
polar in the infrared and the Cartesian in the ultraviolet, and compute the thermodynamic potential.
In the end, we are not convinced from the solutions of the new gap and number equations that the self-consistent
theory is worth the effort. In fact we find results which are not an improvement relative
to those presented in Section VIII at unitarity and the theory has problems in the BEC limit. 

Formally the calculation proceeds in much the same way as in Section IV
with one important difference.
We again write
$\Delta(x) = \Delta_0 + \eta(x)$
where $\Delta_0$ is a real number, which is the $(\xx,\tau)$-independent part of $\Delta(x)$,
and $\eta(x)$ are the complex fluctuations about it. We call this the Cartesian representation
of fluctuations to be contrasted with the polar representation to be introduced below. 
The main difference with Section IV is this: Here $\Delta_0$ does \emph{not} follow the mean field gap equation and its value will be determined only
\emph{after} including the effect of fluctuations, as explained in detail below.
In this sense, the $\eta$'s are \emph{not} the fluctuations about the mean field saddle point,
but rather about a uniform static value $\Delta_0$ which will be determined self-consistently 
\emph{after} integrating out the fluctuations. 
  
We again find that to order $\eta^2$ we get the action
$S_{\Delta} = S_0 + S_g + \ldots$
where $S_0$ has the mean-field-like form (\ref{action_0})
even though $\Delta_0$ is \emph{not} set to its mean field value.
The Gaussian piece too has the same form as (\ref{cartesian}).
We emphasize that there is \emph{no linear term} in
$\eta$ in eq.~(\ref{gaussian_action}), despite the fact that we are not expanding around a saddle point.
The reason for the absence of a linear term is that such a term would be proportional to 
$\eta(q=0)$. However, $\eta(q=0) \equiv 0$ since the uniform ($\qq = 0$), static ($iq_l = 0$) piece of
of $\Delta(q)$ is described by the (as yet undetermined) $\Delta_0$. 

Integrating out the Gaussian fluctuations, we obtain
\bqa
Z \simeq \int d\Delta_0  \int D\eta D\eta^\dagger \exp\left(-S_0 - S_g\right) \\
=\int d\Delta_0 \exp\left[-S_{\rm eff}(\Delta_0)\right]
\label{Z_actionM2}
\eqa
with the effective action
\beq
S_{\rm eff} = S_0 + (1/2) \sum_{\qq,iq_l} \ln {\rm Det} \MM (q).
\label{actionM1}
\eeq
where $S_0$ was defined in eq.~(\ref{action_0}){\bf~\cite{note on Gaussian 2}}.
Using the convergence factors described in Section IV
and Appendix B we get the final result 
\bqa
\no {S_{\rm eff}\over\beta} = \frac{\Delta_0^2}{g} - {1 \over \beta}\sum_{\kk,ik_n} {\rm tr} \ln \gr_0^{-1}(k)~~~~~~~~ \\ 
+ \frac{1}{2\beta} \sum_{\qq,iq_l} \ln \left[{\MM_{11}(q) \over \MM_{22}(q)} {\rm Det} \MM (q)\right]e^{iq_l 0^+}.
\label{actionM2}
\eqa
The parameters $\dl$ and $\mu$ are then fixed by solving the gap equation 
given by 
\beq
\delta S_{\rm eff}/\delta \dl=\delta S_0/\delta \dl +\delta S_g/\delta \dl=0
\label{gap_sc}
\eeq
and the number equation given by
\beq
n=-\partial \Omega /\partial \mu=-\partial[ S_0/\beta]/\partial \mu -\partial [S_g/\beta]/\partial\mu
\label{no_sc}
\eeq

The theory developed above has a serious problem:
there is no zero energy Goldstone mode in the system. To see this 
recall eq.~(\ref{detM_00}) for ${\rm Det}\MM(0,0)$ (which continues
to be valid here). We immediately see ${\rm Det} M(\qq=0,\omega=0)\neq 0$ unless 
$1/g=-\sum_{\bf k} {\rm Det} \gr$. The last condition is however the \emph{mean field} gap equation which 
is \emph{not} satisfied by solutions $\dl$ and $\mu$ of 
the self consistent gap equation (\ref{gap_sc}) and the 
number equation (\ref{no_sc}). We thus see that in the Cartesian representation, 
the Goldstone mode is lost as soon as one moves away from the 
mean field saddle point.

\subsection{Amplitude and Phase Fluctuations in the Self-Consistent Theory}

How can we restore the gapless Goldstone mode in a self-consistent 
Gaussian calculation?
This can be achieved by using a \textit{polar} representation for
the fluctuations in terms of amplitude and phase:
\beq
\Delta(x)=\Do[1+\lambda(x)]e^{i\theta(x)}
\label{polar} 
\eeq
in place of the Cartesian representation $\Delta(x) = \Delta_0 + \eta(x)$ 
used above.
We will first show that the phase excitations $\theta(x)$ are necessarily gapless in 
the long-wavelength limit, 
even when the saddle point shifts away from the mean-field value. 
However, we will find that there is a price to
pay for obtaining the correct low-energy, small-$|\qq|$ physics. 
The high-energy, large-$|\qq|$ behavior of the amplitude-phase fluctuation 
propagator has unphysical properties, and finally we will be forced to an interpolation 
scheme between the polar representation
at low energies and the Cartesian representation at high energies.  

Working with the amplitude $\lambda(x)$ and phase $\theta(x)$ we get
\beq
Z = \int d\Delta_0 \, \int \, D\lambda D\theta \, J \, \exp(-S_{\Delta_0, \lambda, \theta}). 
\label{Z_amplitude_phase}
\eeq
A detailed derivation of the results stated here is given in Appendix F. 
As shown there, the action is the sum of two terms
\beq
S_{\Do,\lambda,\theta} = S_0 + \widetilde{S}_g,
\label{S_amplitude_phase}
\eeq
where $S_0$ defined in eq.~(\ref{action_0}) has the 
mean-field form, and the Jacobian $J$ of the transformation is approximated by
\beq
J = \prod_{{\bf r}, \tau} \Delta_0^2.
\label{S_jacobian}
\eeq
This is the same approximation used in Section IV A of ~\cite{Paramekanti}.  We will see later that the contribution from $J$ is exactly canceled by another contribution [see below eq. (\ref{Dmatrix2})].

The Gaussian term $\widetilde{S}_g$ can be written as
\beq
\widetilde{S}_g =\frac{1}{2}\sum_q \left(\lambda^*(q),\theta^*(q)\right)\DD 
\left(
\begin{array}{clrr}%
        \lambda(q) \\
        \theta(q)
\end{array}
\right).
\label{S_gaussian}
\eeq
We use the notation $\widetilde{S}_g$ here to distinguish it from the 
Gaussian action $S_g$ in the Cartesian case (\ref{cartesian}).
The inverse fluctuation propagator $\DD$ is given by
\beq
\no \DD_{11}=\frac{\Do^2}{g}+\frac{\Do^2}{2}\sum_{\kk,ik_n} {\rm tr}\gr_0(k){\sigma_1}\gr_0(k+q)
{\sigma}_1 \hspace{1cm}
\eeq
\bqa
\no \DD_{22} = \frac{q^2}{8m}\sum_{\kk,ik_n} {\rm tr} \gr_0(k){\sigma}_3\ \hspace{3.2cm}
\\
\no \ + \frac{1}{8}{\rm tr}\gr_0(k)(iq_l{\sigma}_3-\delta\xi)\gr_0(k+q)(iq_l{\sigma_3}-\delta\xi) 
\eqa
\bqa
\no \DD_{12}=  \frac{i\Do}{4}\sum_{\kk,ik_n} {\rm tr} \gr_0(k)(iq_l{\sigma}_3-\delta\xi)\gr_0(k+q){\sigma}_1
\\
\DD_{21}= - \DD_{12}\ \hspace{5cm}
\label{Dmatrix1} 
\eqa
\noindent
where the Pauli matrices ${\bf \sigma}_i$ operate in Nambu space and 
$\delta\xi = \xi_{\kk + \qq} - \xi_\kk = \epsilon_{\kk + \qq} - \epsilon_\kk$.

{\bf Infrared behavior:}
The long-wavelength, low-energy limit of the amplitude and
phase fluctuations described by eq.~(\ref{Dmatrix1}) have the 
following properties. 
(i) $\DD_{12}(\qq=0,iq_l=0)=0$, so that the amplitude and phase modes
decouple in the $q = 0$ limit.
(ii) The $\DD_{22}(q)|\theta(q)|^2$ term, upon
transforming to space-time, has form 
$\rho_s |\nabla\theta|^2/2 - \kappa(\partial\theta/\partial t)^2 + \ldots$,
where $\rho_s$ is the superfluid density and $\kappa$ the compressibility. 
In particular, we note that $\DD_{22}(\qq=0,iq_l=0)=0$ for any choice of 
$\Do$ and $\mu$, so that the phase mode is gapless in the long wavelength 
limit. Thus Goldstone's theorem is respected even when one moves away from 
from the mean field saddle point, in marked contrast to the case 
of Cartesian fluctuations.

Now, it would seem that fluctuations in the amplitude-phase representation 
appear to solve all our problems. It is tempting to argue
that one can simply integrate out the $\lambda$ and $\theta$ fields in 
eq.~(\ref{Z_amplitude_phase})
and obtain an effective action which is the analog of eq.~(\ref{actionM1}) 
with $\sum_{\qq,iq_l} \ln {\rm Det} \MM (q)$ replaced by 
$\sum_{\qq,iq_l} \ln {\rm Det} \DD (q)$.
However the situation is \emph{not} so simple. As we show next, 
the high-energy behavior of ${\rm Det} \DD (q)$
is such that the required Matsubara sum diverges, and there is no  
analog of the convergence factors in eq.~(\ref{actionM2}). 

{\bf Ultraviolet behavior:}
We find it useful to rewrite the $\DD$-matrix in a
form which permits us to better understand its high energy properties
and also to see its relationship to the $\MM$ matrix used to
describe the fluctuations in the Cartesian representation.
Omitting the rather lengthy algebra involved (which is sketched in Appendix F),
we find that 
\beq
\no 
\DD_{11}= \frac{\Do^2}{g}+\frac{\Do^2}{2}\sum_{\kk,ik_n}
\left[\gr^0_{22}\gr^{0\prime}_{11}+ \gr^0_{11} \gr^{0\prime}_{22}+2\gr^0_{12}\gr^{0\prime}_{12}\right],
\eeq
\beq
\no
\DD_{22}=\frac{\Do^2}{2}\sum_{\kk,ik_n}
\left[\gr^0_{22}\gr^{0\prime}_{11}+\gr^{0\prime}_{22}\gr^0_{11}-2\gr^0_{12}\gr^{0\prime}_{12}-2 {\rm Det}\gr^0\right],
\eeq
\beq
\DD_{12}= -\DD_{21} = \frac{i\Do^2}{2}\sum_{\kk,ik_n}
\left[\gr^0_{22}\gr^{0\prime}_{11}-\gr^0_{11}\gr^{0\prime}_{22}\right],
\label{Dmatrix2}
\eeq
where we have used the notation $\gr^0 = \gr^0(k)$ and 
$\gr^{0\prime} = \gr^0(k+q)$.
Note that both the properties discussed below eq.~(\ref{Dmatrix1}) -- 
the decoupling of the amplitude and phase modes at $q=0$ and the Goldstone
theorem -- are also evident in the new expression for the $\DD$ matrix. 

From the expansion of the order parameter field in eq.(\ref{polar}) 
to linear order, we see that the fluctuations of the order parameter 
are of the form 
$\widetilde{\lambda} = \Do\lambda$ and $\widetilde{\theta} = \Do\theta$.
It is useful to rescale the fluctuation fields to $\widetilde{\lambda}$
and $\widetilde{\theta}$, 
so that $\widetilde{\DD} = \DD/\Do^2$. This leads to a factor of 
$2 \ln \Do$ in the action which exactly cancels the factor $J$ in Eq.~(\ref{S_jacobian}). In this 
form it is also easier to make connection with the Cartesian $\eta$ 
fields, as shown below.  
  
At sufficiently high energy and/or short distance scales, the system 
must ``look normal'' (i.e., non-superfluid) and the natural variables to 
describe the fluctuations are the Cartesian $\eta$'s: 
\beq
\left(
\begin{array}{clrr}%
\eta_q \\ 
\eta^*_{-q}
\end{array}\right)=\frac{1}{\sqrt{2}}\left(
\begin{array}{clrr}%
1 & +i \\
1 & -i
\end{array}\right)\left(
\begin{array}{clrr}%
\widetilde{\lambda}_q \\
\widetilde{\theta}_q
\end{array}\right)
=
\WW
\left(
\begin{array}{clrr}%
\widetilde{\lambda}_q \\
\widetilde{\theta}_q
\end{array}\right)\label{cartesian from polar}
\eeq
In this basis the matrix $\widetilde{\DD} \to \LL = \WW \widetilde{\DD} \WW^\dagger$ where 
\beq
\no \LL_{11}=\frac{1}{2}(\widetilde{\DD}_{22}+\widetilde{\DD}_{11}-2i\widetilde{\DD}_{12})=\MM_{11}-\frac{X}{2}
\eeq
\beq
\no \LL_{22}=\frac{1}{2}(\widetilde{\DD}_{22}+\widetilde{\DD}_{11}+2i\widetilde{\DD}_{12})=\MM_{22}-\frac{X}{2}
\eeq
\beq
\LL_{12}=\LL_{21}=\frac{1}{2}(\widetilde{\DD}_{11}-\widetilde{\DD}_{22})=\MM_{12}+\frac{X}{2}
\eeq
where the $\MM$ matrix was defined in eqs.~(\ref{m22},\ref{m12}) and 
\beq\label{Xequation}
X = 1/g+ \sum_{k} {\rm Det}\gr_0(k) = 1/g - \sum_\kk 1/(2E_{\kk}).
\eeq
We now see that, insofar as fluctuations about the mean field saddle point are
concerned, the $\LL$ and $\MM$ matrices are identical. This follows from
the fact that $X \equiv 0$ when the saddle point equation has the mean field form.
(This was the case in the calculation described in Sections V through VIII,
even though $\Do$ and $\mu$ did not have their mean-field values.) 

Conversely, if we look at fluctuations about a saddle point defined
by an equation which does not have the mean field form -- which is the case here -- 
then $X \ne 0$ and the inverse fluctuation propagators $\MM$ 
(directly obtained in the Cartesian representation) and $\LL$ 
(obtained by transformation from the polar $\DD$ to the Cartesian 
representation) necessarily differ.
It is only $\LL$, derived from a polar representation, that respects Goldstone's theorem.
The presence of the $X$ factors which ensure the Goldstone mode in the infrared, 
however, spoils the ultraviolet behavior of $\LL$ and prevents one from 
getting a convergent answer for $\sum_{iq_l} \ln {\rm Det}\LL(\qq,iq_l)$. 
The mathematical analysis showing this difficulty is sketched in Appendix E; 
here we give a simple argument which indicates the problem.

After analytic continuation from $iq_l \to \omega + i0^+$ we find that
in the $\omega \to -\infty$ limit, the $\LL$ matrix looks schematically like
$\LL_{22}(\qq,\omega) \sim - {1 / a_S} + i\sqrt{|\omega|} - X/2 + \ldots$ and
$\LL_{11}(\qq,\omega) \sim - {1 / a_S} + \sqrt{|\omega|} -X/2 + \ldots$ and ${\bf L}_{12} ({\bf q}, \omega) \sim |\omega|^{-3/2} + X/2$.
(We omit multiplicative constants here in various terms and simply focus on
their dependence on $a_S, \omega$ and $X$).
The presence of the $X/2$ factor in the ${\bf L}_{12}$ fundamentally changes its asymptotic behavior from the $|\omega|^{-3/2}$ in the ${\bf M}$ matrix to a constant $X/2$.  This leads to convergence problems discussed in Appendix F.
  
\subsection{Results of the Self-Consistent Calculation}

\begin{figure}[t!]
\includegraphics[width=3in]{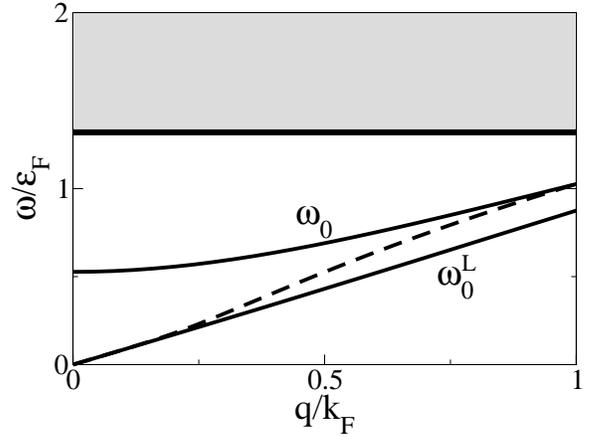}
\caption{Spectra used in the self-consistent calculation of Gaussian fluctuations.  As explained in the text, the method requires to modify the contribution of the poles of the fluctuation matrix at low energies, from that of the cartesian representation to that of the amplitude-phase representation.  The interpolated values used in the calculation are included in the dashed line.}
\label{interpolated.fig}
\end{figure}

We resolve the problem described above
by choosing a scheme that interpolates between the polar representation
in the infrared and the Cartesian representation in the ultraviolet.
We define an ``interpolating'' collective mode energy 
$\omega_0^I(\qq) = \omega_0(\qq) + f(\qq) (\omega_0^L(\qq) - \omega_0(\qq))$
where we choose $f(|\qq|) = 1 - \exp\left(1 - q_c/|\qq|\right)$ for $|\qq| < q_c$ and
$f(|\qq|) = 0$ for $|\qq| \ge q_c$, with $q_c = k_F$ (the only scale at unitarity). 
This formula goes smoothly from 
$\omega_0^I \simeq \omega_0^L$, the polar result at small $q$ to
$\omega_0^I \simeq \omega_0$, the Cartesian result at large $q$,
as shown in Fig.~\ref{interpolated.fig}.
Since the continuum contributions occur at high enough energies (at unitarity) we have left these (Cartesian representation)
contributions untouched.
Operationally, we implement this by adding the following term into the thermodynamical potential
\bqa
\delta \Omega_{sc} &=& {1\over 2} \sum_\qq f(\qq)\, \big [ \omega_0^L(q) - \omega_0(\qq) -\omega_{22}^L(\qq) + \omega_{22}(\qq)\no \\
& & +\omega_{11}^L (\qq)-\omega_{11}(\qq) \big ].
\eqa

Using this method and solving the modified gap and number equations (\ref{gap_sc}, \ref{no_sc}) we obtained that at unitarity the chemical potential is $\mu_{sc} = 0.35\, \epsilon_f$ with a gap of $\Delta_{0, sc} = 0.68 \, \epsilon_f$.  These values compare rather unfavorably with the quantum Monte Carlo values as well as the experimentally measured values. 

We next show that the self-consistent calculation has serious problems in the BEC limit:
the effective interaction between the bosons is attractive and the system is thermodynamically unstable!
Clearly this is an artifact of the modified gap equation. 
 In parallel with the analysis
in Section VII we can show that, just as in eq.~(\ref{Omega_bec}),
$\Omega_g \simeq-\alpha(2m)^{3/2} \dl^4/|\mu|^{3/2}/(256 \pi)$. Although the values of $\dl$ and $|\mu|$ will change because now
we are using a modified gap equation, the dimensionless constant $\alpha$ remains the same as before. As shown in 
Appendix E, it is given by $\alpha = 2.61$.
From the new gap equation we can show that eq.~(\ref{gp1}) is now modified to
\beq
\frac{1}{ a_S}=\sqrt{2m|\mu|}\left[1+ (1- \alpha)\frac{\dl^2}{16|\mu|^2}\right],
\label{gp2}
\eeq
from which we find $\mu=-1/(2ma_S^2)+\delta\mu$ with
\beq
\delta\mu = (1-\alpha){ma_S^2\dl^2 \over 4} = -1.61 {ma_S^2\dl^2 \over 4}.
\label{delta_mu_2}
\eeq
Unlike the result (\ref{delta_mu}) for $\delta\mu$ in Section VII,
we find that the self-consistent calculation yields $\delta\mu < 0$
in the BEC limit. A reduction in $\mu$ relative to the non-interacting boson value
is equivalent to an effective attraction between the bosons or a negative compressibility.  

\section{X. Relation to Other Approaches}

We now turn to a discussion of the relation of our work to that of other authors.
First, the idea of writing the ground state energy density of a many-body system
in terms of the zero-point motion of collective excitations (plasmons) goes back to 
early work on the electron gas using the ``Random Phase Approximation'' (RPA) \cite {rpa}.
The RPA was generalized to the BCS superfluid in the early work of Anderson \cite{Anderson}, where the 
collective mode spectrum and its modification by long-range Coulomb interactions was discussed, but 
the question of the ground state energy density was not fully addressed as far as we can see.
The inclusion of thermally populated collective excitations was central to the Nozieres-Schmitt-Rink
theory of $T_c$ in the BCS-BEC crossover \cite{nozieres,sademelo}. In fact that was the dominant contribution
on the BEC side of the crossover. The difference here is that we are looking at quantum corrections
about the broken symmetry state where we have to deal with matrix propagators.

Several recent works introduce a small parameter by hand; either by expanding in dimensionality around four or
two dimensions \cite{son} or by introducing a large number $2N$ of fermion flavors with a Sp($2N$)-invariant Hamiltonian \cite{Sachdev,Leo}.
{Our self-consistent calculation in Section IX is closely related to the ``$1/N$ expansion'' approach. At zeroth order in 
$1/N$ one obtains the mean field results and first order in $1/N$ gives the RPA or Gaussian fluctuations.  
The saddle point is then recalculated to lowest order in a $1/N$ expansion, with   
changes in the gap and chemical potential from their MF values
obtained perturbatively in $1/N$, which is treated as a small parameter. In practice the calculation is done to first order
in $1/N$ (although in principle it could be done to higher order) and $N$ is set equal to unity at the end.
On the other hand, we keep $N=1$ throughout the self-consistent calculation. Thus the actual values of the $\dl$ and $\mu$ obtained
at unitarity, for instance, are quite different in our approach and in the large $N$ approach, even though if one
was to set $N=1$ throughout the equations would look the same. 
One has to be rather careful about how various physical quantities are calculated in the $1/N$ expansion.
For example, in the BEC limit we can show that $\mu_b = (4\pi n_b/m_b) \,2a_S(1-\alpha/N)$ which is negative for $N=1$ and
would lead to a negative bulk modulus $\partial\mu_b /\partial n_b$. 
On the other hand, the more natural quantity to compute in the grand canonical ensemble is the compressibility $\partial n_b /  \partial \mu_b$
and this is proportional to $1/a_b = (1 + \alpha/N)/(2a_S)$ which is found to be positive even when $N$ is set to unity.}

There have been several diagrammatic and field-theoretical approaches to the crossover problem \cite{Hu-Liu,Haussman,Levin,Ohashi-Griffin,Pieri-Strinati,Diehl}.
Our results in Sections V are essentially the same as the diagrammatic approach of
Hu, Liu and Drummond \cite{Hu-Liu}, although the derivations are somewhat different. In particular,
in the diagrammatic approach the form of the gap equation was unchanged for convenience and only
the thermodynamic potential was altered. In our functional integral framework we can justify this as a 
natural approximation, and, in Section IX we go beyond this approximation and discuss the problems
of  self-consistently including the feedback of Gaussian fluctuations in the gap equation.

The problems that we uncover in the self-consistent approach give insight into the 
conserving approximation scheme used by Haussman and collaborators~\cite{Haussman}.
They too find that, within their approach, as soon as one changes the gap equation from
its mean-field-like form, one has problems with Goldstone's theorem.
They fix this problem by simply redefining the scattering length in an ad-hoc manner
to impose Goldstone's theorem.  Our approach is fundamentally different, as it is based on the observation that the Goldstone mode 
is associated with the presence of soft phase modes in the superfluid phase, amenable to an amplitude-phase decomposition.  

\section{XI. Conclusions}

To conclude, we have studied in this paper the BCS-BEC crossover in an attractive Fermi gas at $T=0$
which is relevant to experiments on ultracold gases with a wide Feshbach resonance. We have
gone beyond the mean field approximation and included the effects of quantum fluctuations at the 
Gaussian level. There is no small parameter which controls this calculation, as we have \emph{not}
introduced a parameter like dimensionality $(4 - \epsilon)$ or number of Fermion species $2N$.
Instead we have attempted to see whether there is an approximation scheme
which can capture the known physics in both the BCS and BEC limits and in addition interpolate
between them through unitarity. 

In summary:
\hfill\break
(1) We include the effect of quantum fluctuations which go beyond mean field theory using a functional
integral approach at $T=0$. We find that at the Gaussian level these fluctuations are the zero-point
motion of the collective modes and the virtual scattering of fermionic quasiparticles.
\hfill\break
(2) In the BCS limit, the virtual scattering of quasiparticles dominates the Gaussian
correction and leads to Fermi-liquid corrections to the ground state properties. 
\hfill\break
(3) In contrast, in the BEC limit the zero-point oscillations dominate the correction term. We
can get an approximate understanding of the renormalization of the effective repulsion between 
molecular bosons and recover the Lee-Yang form for the quantum depletion.
\hfill\break
(4) At unitarity we find that both collective modes and quasiparticle scattering contribute to the thermodynamic potential.
Our results are in good agreement with both quantum Monte Carlo and experimental results.
\hfill\break
(5) We discuss in Section IX the problems of self-consistently including the feedback of fluctuations into the
gap equation. Although the problem of imposing a gapless Goldstone mode is solved by going to the
amplitude-phase representation for the fluctuations, there are still some unsatisfactory aspects
to the calculation. One is the somewhat ad-hoc manner in which the ultraviolet divergences have to be regulated by 
interpolating between the polar and cartesian representations. The results are not quantitatively superior
to the simpler approach at unitarity and there is the further problem of thermodynamic instability in
the BEC limit. In conclusion, we feel it is best to not modify the gap equation by feeding back the
Gaussian fluctuations and to stick to the simpler set of equations dealt with in Sections V through VIII.

\section{Acknowledgments}
The authors would like to thank Matthew Fisher, Parag Ghosh, Jason Ho, Giuliano Orso and Subir Sachdev 
for very useful discussions.
 
\bigskip

\section{Appendix A: Mean Field Theory}

We review the derivation of the mean field gap and number equations with
special attention to convergence factors, which will play a central role
in a more complicated setting later on (see Appendix B).

The saddle-point equation $\delta S_0 / {\delta\Delta_0} =0$,
with $S_0$ given by eq.~(\ref{action_0}), leads to 
the MF gap equation
\beq
\frac{2\dl}{g}={1 \over \beta}\sum_{\kk,ik_n} {\rm Tr}\,\gr_0(k)\frac{\partial\gr_0^{-1}(k)}{\partial \dl}.
\eeq
The Nambu Green's function 
\beq\gr_0(k)=\frac{1}{(ik_n)^2-E_{\kk}^2}
\left(
\begin{array}{clrr}%
        ik_n+\xi_{\kk} &  \ -\Delta_0 \\
        -\Delta_0  & ik_n-\xi_k
\end{array}
\right)
\label{nambuG2}
\eeq
is the inverse of $\gr_0^{-1}(k)$ defined in eq.~(\ref{nambuG0}).
Doing the Matsubara sum we get
\beq
\frac{1}{g} = -{1\over \beta} \,\sum_{\kk,ik_n} \frac{1}{(ik_n)^2-E_{\kk}^2}
= \sum_{\kk}\frac{1-2f(E_{\kk})}{2E_{\kk}}.
\eeq
To obtain the final result (\ref{mf_gap_eq}), we set
the Fermi function $f(E_{\kk})=0$ at $T=0$ in the equation above, and use 
eq.~(\ref{scattering_length}) to take the infinite $\Lambda$ limit and
eliminate the coupling $g$ in favor of the s-wave scattering length $a_S$.
 
Evaluating $n = - \partial\Omega_0 / {\partial\mu}$ with $\Omega_0 = S_0/\beta$ 
we get  
\beq
n={1 \over \beta}\sum_{\kk,ik_n} \left[\gr^0_{11}(k) - \gr^0_{22}(k)\right].
\label{number_eq1}
\eeq 

The Matsubara sum is formally divergent and we must introduce
\emph{convergence factors}. These factors arise because we need to calculate the  
equal time limit of 
$\gr^0_{11}(\kk,\tau)=-\langle T c_{\kk\up}(\tau)\ca_{\kk\up}(0)\rangle$ and 
$\gr^0_{22}(\kk,\tau)=-\langle T \ca_{\kk\dn}(\tau)c_{\kk\dn}(0)\rangle$
to obtain
$n_{\kk\up}= \gr^0_{11}(\kk,\tau\to 0^{-})$
and 
$n_{\kk\dn}= - \gr^0_{22}(\kk,\tau\to 0^{+})$.
We thus rewrite (\ref{number_eq1}) as 
\beq
n={1 \over \beta}\sum_{\kk,ik_n} \left[\gr^0_{11}(k)e^{ik_n0^+}-\gr^0_{22}(k)e^{-ik_n0^+}\right],
\label{number_eq2}
\eeq
and evaluate $\sum_{ik_n}$ as a contour integral in
the complex $z$-plane, with $ik_n \to z$.
The Fermi factors $f(z)= 1/(e^{\beta z}+1)$
ensure convergence for $z \to + \infty$.
For $z\to -\infty$, $e^{z0^+}$ leads to the convergence of the first term  
but the second term is divergent. To convert the offending $e^{-ik_n0^+}$ in (\ref{number_eq2})
to the desired $e^{ik_n0^+}$, we exploit the fact that 
the sum is over both positive and negative $k$ and even
under $k \to -k$, since $\gr^0_{22}(-k) = - \gr^0_{11}(k)$ from eq.~(\ref{nambuG2}).
Thus
\beq
n=2\sum_{\kk,ik_n} \gr^0_{11}(k)e^{ik_n0^+}=
\sum_{\kk}\left[1-\frac{\xi_{\kk}}{E_{\kk}}\tanh (\beta E_{\kk}/2)\right].
\label{number_eq3}
\eeq
The final result going from (\ref{number_eq1}) to (\ref{number_eq3}) could have been simply obtained by 
physical reasoning. The only point of going through the convergence factors in detail here is
that it will streamline the discussion in Appendix B.

\section{Appendix B: Convergence Factors for Bose Matsubara Sums}

In this Appendix, we 
collect useful results for the asymptotic expansion of
$\MM_{ij}$ for large frequencies and show that the integral of the phase $\delta(\qq,\omega)$
of ${\rm Det \MM(\qq,\omega+i0^+)}$ diverges for 
large negative $\omega$. This forces us to introduce convergence factors 
to get a finite answer for the thermodynamic potential, leading us from the formal 
expression (\ref{Omega_1}) to the convergent result (\ref{Omega_2}). 

We use eqs.~(\ref{m22},\ref{m12}) to find the large $q_l$ expansion 
valid for $q_l \gg \max(\Delta_0,|\mu|)$. By neglecting the dependence on $\Delta_0$ and $\mu$
we get
\bqa
&&\no M_{11}(\qq,iq_l) = M_{22}(\qq,-iq_l)\\
&\simeq&- {m\over 4\pi a_S} + \sum_\kk \left ( {1\over iq_l -\epsilon_{\kk + \qq /2} -\epsilon_{\kk - \qq /2}} + {1\over 2\epsilon_\kk} \right )  \no\\
&=& - {m\over 4\pi a_S}+ (m^{3/2}/4\pi) \sqrt{\ep_\qq/2-iq_l},
\eqa
where $\ep_\qq=\qq^2/2m$. For the slightly more restrictive case
when $q_l$ further satisfies $q_l \gg \ep_\qq/2$, we get
\beq
M_{12}=\Delta_0^2m^{3/2}q_l^{-3/2}/(2\sqrt{2}\pi).
\eeq
On the real frequency axis, we are interested in large negative $\omega$
(at $T=0$, the positive frequency contributions go to zero due to the 
Bose occupation factors $n_B(\omega)$).
With $|\omega| \gg \max(\Delta_0,|\mu|)$ we find
\bqa
\no \MM_{11}(\qq,\omega)=-\frac{m}{4\pi a_S}+ \frac{m^{3/2}}{4\pi} \sqrt{|\omega|+\ep_\qq/2} 
+ i {\cal O}(|\omega^{-7/2}|)\\
\MM_{22}(\qq,\omega)=- \frac{m}{4\pi a_S}+ i\frac{m^{3/2}}{4\pi} \sqrt{|\omega|-\ep_\qq/2}.~~~~~~~~~~~~~
\label{M_1_2_real}
\eqa
Here the imaginary part of $\MM_{11}$ comes from just the first term in the sum in 
(\ref{m22}), for which the branch cut appears on the negative real frequency line.
Using the slightly more restrictive condition $|\omega|\gg \ep_\qq/2$
\beq
\MM_{12}(\qq,\omega)= - \frac{\Delta_0^2 m^{3/2}}{4\pi} |\omega|^{-3/2}(1+i)
\eeq
Thus in the limit of large and negative $\omega$ we can neglect $\MM_{12}$ and 
get to the leading order
\beq
 {\rm Det} \MM \simeq - \frac{m^{5/2}}{16\pi^2a_S}\sqrt{|\omega|}+i\frac{m^3}{16\pi^2}|\omega|
\eeq
The phase is given by ${\rm Im} \ln {\rm Det} \MM = \delta \approx \tan^{-1} ( a_S\sqrt{m|\omega|}) \approx \pi/2$ 
as $\omega \rightarrow -\infty$. Thus the Matsubara sum in
eq.~(\ref{Omega_1}) is divergent.
 
We next turn to the derivation of the convergence factors in 
eq.~(\ref{Omega_2}) and how they lead to finite results.
We begin with looking at the sum on Matsubara frequencies for a fixed $\qq$:
\bqa
\no \sum_{iq_l} \ln {\rm Det} \MM(q) \hspace{5cm}
 \\
= \sum_{iq_l}\left[
\ln \MM_{11} + \ln \MM_{22} + \ln \left(1  - {\MM_{12}^2 \over {\MM_{11}\MM_{22}}}\right)\right]
\eqa
and show that the first two terms should be written as
$\sum_{iq_l}\left[\ln \MM_{11}e^{iq_l 0^+} + \ln \MM_{22}e^{-iq_l 0^+}\right]$, while the
third term does not need a convergence factor.
We can rewrite eq.~(\ref{m22}) as
\beq
\MM_{11}(q) = \MM_{22}(-q) = \frac{1}{g}+\sum_{\kk,ik_n} \gr^0_{22}\gr^{0\ \prime}_{11}
\label{m22again}
\eeq
with $\gr^0 = \gr^0(k)$ and $\gr^{0\ \prime} = \gr^0(k+q)$.
We then expand the logarithm in powers of $g$ so that 
$\sum_q \ln \MM_{22}= \sum_q\left[\ln(1/g) + g \sum_k \gr^0_{22}\gr^{0\prime}_{11} + \dots\right]$.
Using the argument given in Appendix A below eq.~(\ref{number_eq1}), we see that
the \emph{equal time limit} requires that 
$\gr_{22}$ carries a factor of $e^{-i-0^+}$ and $\gr'_{11}$ a factor 
of $e^{i(k_n+q_l)0^+}$. We thus see that order by order in $g$ each term in $\ln \MM_{22}$ comes with a 
factor of $e^{-iq_l 0^+}$ and $\ln \MM_{11}$ comes with $e^{+iq_l 0^+}$. 
We also note that, using this prescription,
$\MM_{12}(q)= \MM_{21}(q) = \sum_{\kk,ik_n} \gr^0_{12}\gr^{0\prime}_{12}$
does not acquire a convergence factor and, in fact, none is needed.

The Matsubara sum $\sum_{iq_l}$ is converted to a standard contour integral.
Convergence for $z \to + \infty$ is guaranteed by the Bose function
$n_B(z)= 1/(e^{\beta z} - 1)$. For $z \to -\infty$, convergence is ensured
by converting the problematical factor of $e^{-iq_l 0^+}$
into the convergence factor $e^{iq_l 0^+}$, following the same reasoning as in 
Appendix A. 
Using the fact that the sum is over both positive and negative $q$ and
$\MM_{22}(q) = \MM_{11}(-q)$ (see eq.~(\ref{m22})) we obtain
\bqa
\no \sum_{iq_l} \ln {\rm Det} \MM(q) \hspace{5cm}\\ 
\no =\sum_{iq_l}\left [
2\ln \MM_{11}e^{iq_l 0^+} + \ln \left(1  - {\MM_{12}^2 \over {\MM_{22}\MM_{11}}}\right)\right]\\
=\sum_{iq_l}\ln \left[\frac{\MM_{11}{\rm Det} \MM}{\MM_{22}}\right]e^{iq_l 0^+}~~~~~~~~~~~~~~~~~~~~~~~
\label{lnDetM}
\eqa
which is exactly the expression in eq.~(\ref{Omega_2}).

We finally show explicitly that the Matsubara sum in eq.~(\ref{lnDetM}) 
is convergent. The Matsubara sum in eq. (\ref{lnDetM}) can be written  
as the contour integral 
$\oint_{\cal C} dz/(2\pi i) n_B(z) \ln [\MM_{11}(\qq,z) {\rm Det} \MM(\qq,z)
/\MM_{22}(\qq,z)]$
where ${\cal C}$ runs on either side of the imaginary z axis, enclosing 
it counterclockwise. We distort the contour to run above and below the 
real axis and at $T=0$ obtain for the thermodynamic potential   
\beq
\Omega_g=-\frac{1}{2}\sum_\qq\int_{-\infty}^0 \frac{d\omega}{\pi} [\delta(\qq,\omega)+\delta_{11}(\qq,\omega)-\delta_{22}(\qq,\omega)]
\label{phase}
\eeq
 where $\delta(\qq,\omega) = {\rm Im} \ln {\rm Det} \MM(\qq,\omega+i0^+)$ and $\delta_{11}$ and $\delta_{22}$ are the 
corresponding phases for $\MM_{11}$ and $\MM_{22}$. 

From the leading order expression for $\MM_{11}$ in eq. (\ref{M_1_2_real}), 
we see that for large negative $\omega$, $\delta_{11}\sim |\omega|^{-4}$ 
and hence that term is convergent. To look at $\delta-\delta_{22}$, we 
recognize that this is the phase of ${\rm Det}\MM/\MM_{22}=\MM_{11}-\MM_{12}^2/\MM_{22}$. Now for large negative $\omega$, 
$\MM_{12}^2/M_{22}\sim |\omega|^{-7/2}+i|\omega|^{-4}$ and can be neglected 
in comparison to $\MM_{11}$. Thus the integrand in eq. (\ref{phase}) reduces 
to $2\delta_{11}$ and we get a convergent answer.  

\section{Appendix C: Numerical Evaluation of Bose Matsubara Sums}

While the real frequency representation of eq.~(\ref{Omega_2}) gives physical insight into 
the deviations away from mean field theory, it is numerically simpler to
do the calculation on the imaginary frequency axis. On the real axis one encounters 
principal part singularities analogous to the ones encountered, e.g., in the
normal state calculations of ref.~\cite{engelbrecht_2dfermigas}
but further complicated by the broken symmetry in the superfluid state.

If one wants to use eq.~(\ref{Omega_2}) on the Matsubara axis, one needs to explicitly take 
into account the convergence factor $e^{+iq_l \tau}$ 
and take the limit $\tau \to 0^+$ at the end. Here we outline an alternative
procedure which simplifies the numerics.
Let us begin by looking at a part of $\MM_{11}(\qq,z)$ 
\beq
\no \MM^C_{11}(q) 
=\frac{1}{g}+\sum_\kk \frac{u^2u'^2}{iq_l - E - E'} = \MM^C_{22}(-q)
\label{m11c}
\eeq
which has no singularities (poles, branch cuts) or zeros in the left-half plane
(${\rm Re}\,z < 0$). Since we will use this to get convergent results
we call it $\MM_{11}^C$ and $\MM_{22}^C$.

We may write the $\MM_{11}$ piece of eq.~(\ref{lnDetM}) as
\beq
\sum_{iq_l} 2\ln \MM_{11}e^{iq_l0^+} = 
\sum_{iq_l} 2\left[\ln (\MM_{11}/\MM_{11}^C) + \ln \MM_{11}^C\right]
\eeq
where we drop $e^{iq_l0^+}$ on the right because each term is convergent.
The Matsubara sum of the second term is seen to be zero at $T=0$ by evaluating it as
a standard contour integral and noting that $\ln \MM_{11}^C$ has no singularities in 
the left-half plane. In fact, now we may write the above result in a more
symmetrical form as
$\sum_{iq_l} \left[\ln (\MM_{22}/\MM_{22}^C) + \ln (\MM_{11}/\MM_{11}^C)\right]$
and combine this with the second term of (\ref{lnDetM}) to obtain
\beq
\sum_q \ln {\rm Det} \MM(q) \to  \sum_q \ln {\rm Det} \left[\MM(q) \over \MM^C(q)\right]
\label{lnDetM2}
\eeq
where the matrix $\MM^C(q)$ is a diagonal matrix with the entries $\MM_{22}^C(q)$ and $\MM_{11}^C(q)$.
This expression leads to a rapidly convergent answer, which  
in the $T=0$ limit can be evaluated as an integral along the imaginary axis in the $iq_l \to z=(x+iy)$ plane
with $\beta^{-1}\sum_{iq_l} \to \int dy /(2\pi)$.  

\section{Appendix D:  Fermi liquid corrections for $k_F|a_S| \ll 1$ with $a_S <0$} 

Here we give some details of the argument that shows that the Gaussian corrections to
the thermodynamic potential in the extreme BCS limit of the attractive Fermi gas 
have the same expression as the standard Galitskii and Huang-Lee-Yang theory 
\cite{galitskii,Lee-Yang,Fetter-Walecka} of the repulsive 
Fermi gases with a sign change in $a_S$. First we recall the usual Galitskii theory and discuss
why it is not directly useful for $a_S < 0$. Next, we describe how the
BCS limit results of the superfluid state theory developed in the paper are
related to those of normal state Galitskii theory.

\begin{figure}[t]
\includegraphics[scale=0.4]{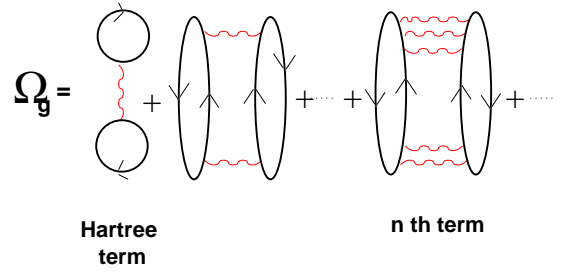}
\caption{Diagrammatic representation of the Gaussian corrections to thermodynamic potential in the BCS limit. The full lines are fermion propagators and the wave lines represent the attractive interaction.  The first diagram corresponds to the Hartree term.}  
\label{lyg}
\end{figure}

In the Galitskii theory of the normal Fermi gas the thermodynamic potential $\Omega$ is
written in terms of the two-particle propagator, which is the sum of particle-particle 
channel ladder-diagrams of Fig.~\ref{lyg}. For a repulsive interaction $V_0$, we find
\beq 
\Omega = \Omega_{\rm free} + 2\sum_{\ell=1}^\infty
(-1)^{\ell}\frac{V_0^\ell}{\ell}\sum_q \left[\sum_k G_0(-k)G_0(k+q)\right]^\ell
\label{galitskii_1}
\eeq
where $G_0$ is the non-interacting Green's function 
$G_0(k)=(ik_n-\ep_\kk+\mu)^{-1}$. To make contact with results of our paper, it is useful to
sum up the series and write it as
\beq 
\Omega = \Omega_{\rm free} + 2\sum_{q} \ln[1 - V_0\sum_k G_0(-k)G_0(k+q)].
\label{galitskii_2} 
\eeq 
The repulsive $V_0$ can then be replaced by $a_S > 0$
in the usual way using $m/4\pi a_S=1/V_0+\sum_\qq 1/2\epsilon_\qq$.

\begin{figure}[t]
\includegraphics[scale=0.4]{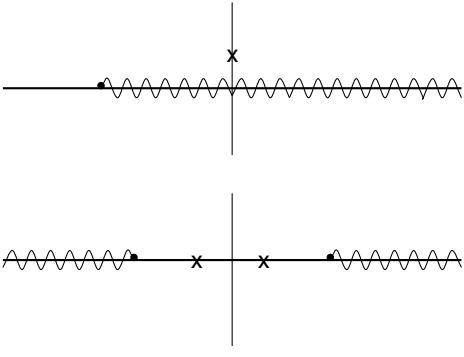}
\caption{Analytic structure of two-particle propagators for the attractive Fermi gas. 
The upper panel correspond to the unstable normal state which has a branch cut for $\omega \ge -2\mu +\epsilon_{\qq}/2$,
representing the continuum of excitations, and a pole on the imaginary axis in the upper-half plane, for small $\qq$, which signals
the BCS pairing instability. The lower panel shows the analytic structure of $\MM^{-1}$ in the
stable superfluid state with poles at $\pm\omega_0(\qq)$, the collective mode frequencies,
and branch cuts beginning at $\pm {\min}\left(E_\kk + E_{\kk+\qq}\right)$.
}
\label{galitskii_analytical}
\end{figure}

It is well-known that the pairing instability of the normal Fermi gas to attractive interactions 
implies that we \emph{cannot} extend the Galitskii calculation directly to the case of
attractive interactions $a_S < 0$. If we were to try and set $V_0 = -g$, the attraction
of eq.~(\ref{hamiltonian}), we would find that, for small $q$, there is a pole on the imaginary axis
in the \emph{upper-half plane} in addition to a branch cut along the real axis
$\omega \ge -2\mu +\epsilon_{\qq}/2$, as shown in the top panel of Fig.~\ref{galitskii_analytical}.
This pole, which occurs at $z \sim + i \epsilon_f \exp{(-1/k_F|a_S|)}$ for $\qq = 0$,
is the signature of the Cooper pairing instability.

The superfluid state calculation for the attractive Fermi gas
discussed in the text of this paper deals with the broken symmetry
saddle point which is stable. The two particle propagator $\MM^{-1}$ in the superfluid state has a very
different analytical structure (lower panel of Fig.~\ref{galitskii_analytical}) compared
with the unstable normal state just discussed (upper panel of Fig.~\ref{galitskii_analytical}).
As described in Section IV below eq.~(\ref{Omega_2}), $\MM^{-1}$ has poles at
the collective modes frequencies and branch cuts corresponding to the gapped two-particle continuum.

In the BCS limit, $\dl \rightarrow 0$ and the contribution of the zero-point motion of the collective modes to $\Omega$
is utterly negligible for phase-space reasons, as discussed in Section VI. Thus the two-particle continuum
dominates the Gaussian correction to $\Omega$.
In the limit $\dl \rightarrow 0$ the branch cut extends over the entire real line, although the appropriate
limit of the $u_k, v_k$ factors shows that the phase shift vanishes for $\omega < -2\mu +\epsilon_{\qq}/2$.
In this sense the continuum contribution to $\Omega$ given by $\int_{-\infty}^{-E_c(\qq)} d\omega \delta(\qq,\omega)$
may be simplified with the lower limit becoming $-2\mu +\epsilon_{\qq}/2$ and the upper limit going to zero.

Now one can check that this continuum contribution is exactly the same 
as the corresponding continuum contribution of the normal state Galitskii theory, changing the sign of $a_S$.
Note that the singular pole piece does \emph{not} show up in this result. 
To see that the $\delta(\qq,\omega)$'s in the two theories are the same, start with the $\MM$ matrix 
of the broken symmetry theory. As $\dl \rightarrow 0$, $M_{12}$ vanishes and 
$\delta(\qq,\omega) \simeq 2 {\rm Im} \ln \MM_{11} =2 {\rm Im} \ln[1-g\sum_k G_0(-k)G_0(k+q)]$, and we have used convergence factor tricks
to obtain well-defined Matsubara sums.
The last expression is the same as the phase shift obtained from
the normal state result above (\ref{galitskii_2}) with $V_0 \to -g$.
We thus conclude that in the $\dl \rightarrow 0$ limit of the superfluid state,
the Gaussian correction to $\Omega$ is the same as the Galitskii and Huang-Lee-Yang result for the
repulsive Fermi gas with sign of $a_S$ changed to $a_S<0$.

\section{Appendix E:  Thermodynamic potential in the BEC limit}

Here we briefly sketch how we arrive at the leading 
order correction to the thermodynamic potential in the BEC limit 
eq.~(\ref{Omega_bec}) starting with the results of Appendix C. In the BEC limit, 
$\mu < 0$ and approaches half the binding energy of the molecules 
$|\mu|=1/(2ma_S^2)$. Thus $\dl/|\mu| \ll 1$ and can be used as an expansion 
parameter. Then one can write $u_\kk^2=1-\dl^2/4\xi_\kk^2$, 
$v_\kk^2= \dl^2/4\xi_\kk^2$ and $E_\kk=\xi_\kk+\dl^2/2\xi_\kk$ with 
$\xi_\kk=\ep_\kk+|\mu|$ to leading order in $\dl/|\mu|$.

 Now writing $\MM_{11}=\MM_{11}^C+\delta\MM_{11}$, where 
$\MM_{11}^C$ is defined in eq. (\ref{m11c}) and 
\beq
\delta\MM_{11}(q)=-\sum_\kk \frac{v^2v'^2}{iq_l+E+E'}
\eeq  
one can easily see that $\delta\MM_{11}\sim \dl^4$ while $\MM_{12}\sim \dl^2$.
 Then, to order $\sim \dl^4$, the expression in eq. (\ref{lnDetM2}) gives 
\beq
\Omega_g=\sum_q {\rm Re} \left(\frac{\delta\MM_{11}}{\MM_{11}^C}\right)-\frac{\MM_{12}^2}{2|\MM_{11}^C|^2}
\eeq
To leading order, 
$\delta \MM_{11}=\dl^4/|\mu|^{7/2}F(\qq/\sqrt{2m |\mu|},q_l/|\mu|)$ 
where $F(\QQ,Q_l)$ is a dimensionless function given by
\bqa
F(\QQ,Q_l) &=& {1\over 16} 
\sum_{\KK} {1\over iQ_l +2 +2K^2+Q^2/2} 
\\ 
& & 
{1\over ((\KK + \QQ/2)^2+1)^2 ((\KK - \QQ/2)^2+1)^2}
\nonumber 
\eqa
Here we use capital letters for dimensionless variables
$\QQ = \qq/\sqrt{2m |\mu|}$ and $Q_l = q_l/|\mu|$.
$\MM_{12}=\dl^2/|\mu|^{3/2} I(\qq/\sqrt{2m |\mu|},q_l/|\mu|)$ 
where $I(\QQ,Q_l)$ is a dimensionless function given by
\bqa
I(\QQ,Q_l) &=& -{1\over 4} 
\sum_{\KK} {2 +2K^2+Q^2/2 \over Q_l^2 +(2 +2K^2+Q^2/2)^2}
\\  
& & {1\over ((\KK + \QQ/2)^2+1) ((\KK - \QQ/2)^2+1)}
\nonumber
\eqa
and $\MM_{11}^C=\sqrt{|\mu|}H(\qq/\sqrt{|\mu|},q_l/|\mu|)$ 
where $H(\QQ,Q_l)$ is a dimensionless function given by
\bqa
H(\QQ,Q_l)&=& {1\over 8\pi} \left (\sqrt{-i{Q}_l/2 +{Q}^2/4+1} - {1 \over  \sqrt{2m} a_S \sqrt{|\mu|} }\right )\nonumber \\
&\approx&  {1 \over 8\pi} \left (\sqrt{-i{Q}_l/2 +{Q}^2/4+1} - 1 \right )
\end{eqnarray}

Putting all these together, we get 
\bqa
{\Omega_g} & = & \frac{(2m)^{\frac{3}{2}}\dl^4}{4\pi^3|\mu|^{3/2}}\int d^3\QQ\int dQ_l 
\nonumber\\ 
& & 
\left\{ {\rm Re}\left[\frac{F(\QQ,Q_l)}{H(\QQ,Q_l)}\right]
-\frac{[I(\QQ,Q_l)]^2}{2\left\vert H(\QQ,Q_l)\right\vert^2} \right\}
\eqa
Numerical evaluation of the integral gives 
${\Omega_g} =- \alpha (2m)^{3/2}\dl^4/|\mu|^{3/2}/(256 \pi)$ with $\alpha=2.61$.

\section{Appendix F: The Amplitude-Phase Action}

Starting with $\Delta(x)=\Do[1+\lambda(x)]e^{i\theta(x)}$ (see eq.~(\ref{polar})) 
we transform to a gauge where $\Delta(x)$ is real. We transform the 
fermion fields
\beq
\widetilde{\psi}(x)=\UU(x)\psi(x)
\eeq 
with
\beq
\UU(x)=\left(
\begin{array}{clrr}%
        e^{-i\theta(x)/2} &   0 \\
        0  & e^{i\theta(x)/2}
\end{array}
\right)
\eeq
so that the action (\ref{action}) now reads
\beq
S_{\widetilde{\psi},\Delta}= \int dx \frac{1}{g}|\Delta(x)|^2-\widetilde{\psi}^{\dagger}(x)\widetilde{\gr}^{-1}(x,x')\widetilde{\psi}(x')
\label{action_1}
\eeq
where
$\widetilde{\gr}^{-1} = \UU \gr \UU^{\dagger}$.

We can now write $\widetilde{\gr}^{-1}=\gr_0^{-1}+\widetilde{\bf K}$, where 
$\gr_0^{-1}$ is the (inverse) Nambu Green's function 
defined by (\ref{nambuG}) in $k=(\kk,ik_n)$-space. 
The matrix $\widetilde{\bf K}$ is 
\bqa
\no \widetilde{\bf K}(x,x')=\left[\Do\lambda(x)\sigma_1
+\frac{i}{2m}\left(\nabla\theta(x)\cdot\nabla+
{1 \over 2}\nabla^2\theta(x)\right)\right.\\
\left.-\left(\frac{i}{2}\partial_{\tau}\theta(x)
+\frac{1}{8m}|\nabla\theta(x)|^2\right)\sigma_3\right]\delta(x-x') \hspace{0.6cm} 
\label{Kmatrix1}
\eqa
whose Fourier transform is
\bqa
\no \widetilde{\bf K}(k',k)=\left[\Do\lambda_q\sigma_1+\frac{i}{2}(iq_l\sigma_3-\delta\xi)\theta_q\right]\delta(k-k'+q)\\
+\frac{1}{8m}\sum_{q_1q_2}q_1\cdot q_2\theta_{q_1}\theta_{q_2}
\sigma_3\delta(k-k'-q_1-q_2)\hspace{0.5cm} 
\label{Kmatrix2}
\eqa
with $\delta\xi=\xi_{\kk+\qq}-\xi_\kk$.

Integrating out the fermion fields $\widetilde{\psi}$ we get the 
functional integral (\ref{Z_amplitude_phase}) with the action 
$S_{\Do,\lambda,\theta} = S_0 + \widetilde{S}_g$ of eq(\ref{S_amplitude_phase}). 
The $S_0$ piece, defined in (\ref{action_0}), comes from the $\gr_0^{-1}$ term; for the $J$ term in (\ref{Z_amplitude_phase}) see (\ref{S_jacobian}).
The Gaussian piece, arising from $\widetilde{\bf K}$, is given by
\bqa
\no \widetilde{S}_g = \frac{\dl^2}{g}\lambda_q\lambda_{-q}- {\rm Tr}\gr_0(k)\widetilde{\bf K}(k,k)~~~~~~~~~~~~~~ \hspace{1.5cm}\\
 +\frac{1}{2}Tr \gr_0(k)\widetilde{\bf K}(k,k+q)\gr_0(k+q)\widetilde{\bf K}(k+q,k)~~~ 
\label{S_gaussian1}
\eqa

The Gaussian action of (\ref{S_gaussian}) follows immediately from (\ref{Kmatrix2}) and (\ref{S_gaussian1}),
with the $\DD$ matrix given by (\ref{Dmatrix1}). Our next task is to derive the equivalent expression for
the $\DD$ matrix (\ref{Dmatrix2}) which is written purely in terms of $\gr^0$, without any $iq_l\sigma_3 - \delta\xi$
factors. The case of $\DD_{11}$ is simple; there are no such factors to begin with  
and we only need to evaluate the Nambu trace in (\ref{Dmatrix1}) to obtain (\ref{Dmatrix2}).
In what follows, we use
the notation $\gr=\gr_0(k)$ and $\gr'=\gr_0(k+q)$, and drop the subscript 
$0$ for notational convenience.

In order to write $\DD_{12}$ and $\DD_{22}$ in terms of the Green's functions, 
one needs to express the vertex $iq_l\sigma_3-\delta\xi$ in terms of matrix 
elements of $\gr^{-1}$. It is easy to see that the vertex can be written 
as 
\begin{center}
\begin{math}
iq_l\sigma_3-\delta\xi= \VV=\left(
\begin{array}{clrr}%
        \delta_{11} &   0 \\
        0  &  -\delta_{22}
\end{array}
\right)
\end{math}
\end{center}
where $\delta_{11}=\gr'^{-1}_{11}-\gr^{-1}_{11}$ and $\delta_{22}=\gr'^{-1}_{22}-\gr^{-1}_{22}$.
We will also use the following identities:
\begin{eqnarray}
\no \gr_{22}\gr^{-1}_{22}=1-\Do \gr_{12} \,&,& \, \gr_{11}\gr^{-1}_{11}=1-\Do \gr_{12}\\
\gr_{12}\gr^{-1}_{22}=-\Do\gr_{11}  \,&,& \,  \gr_{12}\gr^{-1}_{11}=-\Do \gr_{22} \label{identity}
\end{eqnarray}

By definition we have 
\bqa
\no \DD_{12}=\Do\frac{i}{4}Tr \gr_0(k+q)\VV\gr_0(k)\sigma_1~~~~~~\\
\no =\frac{i\Do}{4}\sum_k \delta_{11}(\gr_{11}\gr'_{12}+\gr_{12}\gr'_{11})\\
-\delta_{22}(\gr_{12}\gr'_{22}+\gr_{22}\gr'_{12})
\eqa
Now using the identities of eq. (\ref{identity}), we get
$
\no \sum_k \delta_{11}(\gr_{11}\gr'_{12}+\gr_{12}\gr'_{11})= 
\sum_k\dl(\gr_{22}\gr'_{11}-\gr_{11}\gr'_{22})
$
and a similar result holds for the $\delta_{22}$ piece.   
Adding both terms we get $\DD_{12}$ of eq.~(\ref{Dmatrix2}).
For $\DD_{22}$, one can write
\beq
\DD_{22}=\frac{q^2}{8m}\sum_k (\gr_{11}-\gr_{22})+\frac{1}{8}Tr \gr_0(k)\VV\gr_0(k+q)\VV
\label{D11}
\eeq
The second term above can be written as
\begin{center}
\begin{math}
\frac{1}{8}\sum_k \delta^2_{11}\gr_{11}\gr'_{11}+\delta_{22}^2\gr_{22}\gr'_{22}
-2\delta_{11}\delta_{11}\gr_{12}\gr'_{12}~~~~~~~~~~~~~~~~~~
\end{math}
\end{center}
Using the identities in eq. (\ref{identity}), we get
\bqa
\no \sum_k \gr_{11}\gr'_{11}\delta_{11}^2 =
\Do^2\sum_k \gr_{11}\gr'_{22}+\gr'_{11}\gr_{22}-2\gr_{12}\gr'_{12}\\
\no+2\Do\sum_k(\gr_{12}+\gr'_{12})~~~~~~~~~~~~~~~~~~~~~~~\\
\no+\sum_k \gr_{11}\gr'^{-1}_{11}+\gr'_{11}\gr^{-1}_{11}-2~~~~~~~~~~~~
\eqa
and a similar result holds for the $\delta_{22}^2$ piece.
For the last term we get
\bqa
\no-2\sum_k \gr_{12}\gr'_{12}\delta_{11}\delta_{22} 
= 2\Do^2\sum_k \left[\gr_{11}\gr'_{22}+\gr'_{11}\gr_{22}\right.\\
\no \left. -2\gr_{12}\gr'_{12}\right]
+2\Do\sum_k(\gr_{12}+\gr'_{12})
\eqa
Adding all the terms we get
\bqa
\no \frac{\Do^2}{2}\sum_k \gr_{22}\gr'_{11}+\gr'_{22}\gr_{11}-2\gr_{12}\gr'_{12}+ \frac{3\Do}{2}\sum_k\gr_{12}~~~~~~\\
\no+\frac{1}{8}\sum_k \gr_{22}\gr'^{-1}_{22}+\gr'_{22}\gr^{-1}_{22}+
\gr_{11}\gr'^{-1}_{11}+\gr'_{11}\gr^{-1}_{11}-4~~~~~~
\label{d11_int}
\eqa
where we have used $\sum_k\gr_{12}=\sum_k\gr'_{12}$ .
We now use $\gr'^{-1}_{11}=\gr^{-1}_{11}+iq_l-\delta\xi$ and 
$\gr'^{-1}_{22}=\gr^{-1}_{22}+iq_l+\delta\xi$  to write the last term as
$
-\frac{1}{2}\Do\sum_k \gr_{12}+\frac{iq_l}{2}(\gr_{11}+
\gr_{22}) -\frac{\delta\xi}{4}(\gr_{11}-\gr_{22} - \gr'_{11}+
\gr'_{22})$,
where we have used $\sum_k \gr_{ij}=\sum_k\gr'_{ij}$ to write this form.
Now using proper convergence factors $\sum_k \gr_{11}+\gr_{22}=0$ and so the 
terms multiplying $iq_l$ vanishes .
In the last term, replace ${\bf k}  \rightarrow \kk  + \qq$ to show that this term is proportional to $\qq^2/2m$ and it actually 
exactly cancels the similar term in $\DD_{22}$ coming from the 
$(\qq^2/2m){\rm Tr}\gr\sigma_3$ piece. Now $\sum_k\gr_{12}=-\Do\sum_k {\rm Det}\gr$, and 
so combining everything we get the result for $\DD_{22}$ in (\ref{Dmatrix2}).
\beq
\DD_{22}=\frac{\Do^2}{2}\sum_k \gr_{11}\gr'_{22}+\gr'_{11}\gr_{22}-2\gr_{12}\gr'_{12}-2 Det \gr
\eeq
Going to the rescaled basis $(\widetilde{\lambda}, \widetilde{\theta})$, 
we then have 
\bqa
\no 
\widetilde{\DD}_{11}= \frac{1}{g}+\frac{1}{2}\sum_{k}
\left[\gr_{22}\gr_{11}'+ \gr_{11} \gr_{22}'+2\gr_{12}\gr_{12}'\right]
~~~~~~~\\
\no
\widetilde{\DD}_{22}=\frac{1}{2}\sum_k
\left[\gr_{22}\gr'_{11}+\gr'_{22}\gr_{11}-2\gr_{12}\gr'_{12}-2 {\rm Det}\gr\right]\\
\widetilde{\DD}_{12}=\frac{i}{2}\sum_k
\left[\gr_{22}\gr'_{11}-\gr_{11}\gr'_{22}\right]~~~~~~~~~~~~~~~~~~~~~~~~~~~~
\eqa

Just as in the case of the static Saddle Point number equation, one runs into 
formally divergent quantities in evaluating the $q$ sum to get the action.
To fix these, one has to regularize using proper convergence factors. The 
$D$ basis is not the basis of choice for fixing the convergence factors.
Instead of the amplitude $\widetilde{\lambda}$ and the phase 
$\widetilde{\theta}$ we can work with the complex fluctuation fields defined in (\ref{cartesian from polar}).
In this basis the matrix $\DD$ is transformed to 
\beq
\LL = \WW \widetilde{\DD} \WW^{\dagger}
\eeq

Since this is an unitary transform ${\rm Det} \widetilde{\DD}= {\rm Det} \LL$.
Then we get
\bqa
\no L_{11}=\frac{1}{g}+\sum_k\gr_{22}\gr'_{11}-\frac{X}{2}=\MM_{11}-\frac{X}{2}\\
L_{12}=\sum_k\gr_{12}\gr'_{12}+\frac{X}{2}=\MM_{12}+\frac{X}{2}\\
\no L_{22}=\frac{1}{g}+\sum_k\gr_{11}\gr'_{22}-\frac{X}{2}=\MM_{22}-\frac{X}{2}
\eqa
where $X= 1/g+\sum_k Det \gr = 1/g-\sum_\kk 1/(2E_{\kk})$ is the LHS of the 
mean field gap equation and 
$\LL_{12}=\LL_{21}$.

Now we can fix convergence factors with $\ln \LL_{11}$ carrying a convergence 
factor of $e^{+iq_l0^+}$ and $\ln \LL_{22}$ carrying a convergence factor 
of $e^{-iq_l0^+}$. This can be seen by expanding the log and remembering 
$\gr_{22}$ carries a factor of $e^{-ik_n0^+}$, $\gr'_{11}$ carries a factor 
of $e^{+i(k_n+q_l)0^+}$ and so on. The reasons for the convergence factors 
are related to taking the correct equal time limit and is discussed in 
 in detail in Appendices A and B.

Remembering $\LL_{11}(-q)=\LL_{22}(q)$ one can take out $\ln \LL_{11}+ \ln \LL_{22}$ 
from the $\ln {\rm Det} \LL$ and then convert $\ln \LL_{22}$ to $\ln \LL_{11}$ using
$q\rightarrow -q$. Now one can convert the Matsubara sums to real frequency 
integrals which are convergent. The resulting action is

\beq
S_g=\frac{1}{2}\sum_q \ln \left(\frac{\LL_{11}}{\LL_{22}} {\rm Det}L\right)
\eeq
One can then follow the asymptotic forms of the $\MM$ matrix derived in 
Appendix E to get the large energy-short wavelength behavior of the $\LL$ 
matrix. The asymptotic forms of $\LL_{11}$ and $\LL_{22}$ are the same as 
that of $\MM_{11}$ and $\MM_{22}$, with $m/4\pi a_S$ replaced by 
$m/4\pi a_S-X/2$. However the presence of the $X/2$ factor in $\LL_{12}$ 
fundamentally changes its asymptotic behavior from $\omega^{-3/2}$ for 
the $\MM$ matrix to a constant ($X/2$). We can thus no longer neglect the 
$\LL_{12}$ terms in the high frequency limit and this leads to a divergent 
answer.

\section{Appendix G:  Dilute Bose gas}

In this appendix we show how the method of Gaussian fluctuations yields the correct answers in a somewhat
different problem, that of a dilute {\em Bose} gas with repulsive interactions. Although the
results are standard \cite{Fetter-Walecka}, the method used here parallels that used in
our paper, and serves to illustrate several technical points including: 
(1) the role of convergence factors, (2) retaining the mean field form of the saddle point equation and 
including quantum fluctuations in the thermodynamic potential, and 
(3) taking into account the $\mu$-dependence of the the saddle point in the number density equation. 

The Hamiltonian for a repulsive ($g_0 > 0$) Bose gas is
\bqa
H&=&\int d^3 x \Phi^*(x)(-\nabla^2/2M -\mu)\Phi(x)\no \\
& & +{g_0\over2} \Phi^*(x)\Phi^*(x)\Phi(x)\Phi(x)
\eqa
Writing $\Phi(x)=\Phi_0+\zeta(x)$ the action is $S = S_0 + S_g + \ldots$
where
\beq
S_0 = \beta\left(-\mu \Phi_0^2+{g_0\over2} \Phi_0^4\right)
\label{bose_mf_action}
\eeq
and the Gaussian part is given by
\beq
S_g = {1 \over 2}\sum_{\qq,iq_l} \left( \zeta^*(q),\zeta(-q) \right)
\A(q)
\left(
\begin{array}{clrr}%
        \zeta(q) \\       
        \zeta^*(-q)
\end{array}
\right).
\label{bose_gaussian}
\eeq
Here $\A_{11}(q) = \A_{22}(-q) = -iq_l + \epsilon_\qq - \mu + 2g_0\Phi_0^2$,
with $\epsilon_\qq = |\qq|^2/2M$, and $\A_{12}(q) = \A_{21}(-q) = g_0\Phi_0^2$.
Integrating out the $\zeta$ fields we 
get the thermodynamic potential 
\beq
\Omega \simeq \Omega_0 + (1/2\beta) \sum_{\qq,iq_l} \ln {\rm Det} \A (q),
\eeq
where $\Omega_0= S_0/\beta$. The Matsubara sum in the
Gaussian piece is ill-defined. We write $\ln {\rm Det} \A (q) = 
\ln \A_{11} + \ln \A_{22} + \ln \left(1  - {\A_{12}^2 /{\A_{22}\A_{11}}}\right)$,
introduce convergence factors of $\exp(iq_l0^+)$ with the $\A_{11}$ term (associated with
$\zeta^*\zeta$), and
$\exp(-iq_l0^+)$ with the $\A_{22}$ term (corresponding to $\zeta\zeta^*$) and
use $q \to -q$ to write the $\A_{22}$ piece in terms of $\A_{11}$.
At $T=0$ the sum $ \beta^{-1}\sum_{iq_l} 2 \ln \A_{11} \exp(iq_l0^+)$ vanishes
by contour integral methods since the integrand has no singularities in the
left-half plane. The remaining sum can be explicitly done by contour methods to obtain
\beq\label{Omega Bosons}
\Omega = \Omega_0 + {1\over 2}\sum_\qq \left(E_\qq-\epsilon_\qq+\mu-2g_0\Phi_0^2\right)
\eeq
where $E_\qq=\sqrt{(\epsilon_\qq-\mu+2g_0\Phi_0^2)^2-g_0^2\Phi_0^4}$ is the Bogoliubov
dispersion. The quantum fluctuations are clearly seen to have the form of zero
point motion of the collective modes $E_\qq/2$ with a ``convergence factor'' subtraction
which eliminates the ultraviolet divergence by canceling out the
contribution of the quadratic part of the Bogoliubov spectrum at large $q$. 

The uniform, static saddle point is determined by 
$\delta S_0 / \delta\Phi_0 = 0$, so that
\beq
\Phi_0^2 = \mu/g_0
\label{mf}
\eeq
{and this condition is again needed in order to satisfy that the excitation spectrum is gapless}.  We use $\left( \partial\Omega/\partial\mu \right)=-N$ to determine $\mu$.
In evaluating the thermodynamic derivative
we {\it cannot} treat $\Phi_0$ as a constant, and must
keep track of the $\mu$-dependence of $\Phi_0$ in eq.~(\ref{mf}). 
We thus get
$\Omega= -{\mu^2 / 2g_0} + {1\over2}\sum_\qq \left(E_\qq-\epsilon_\qq-\mu\right)$
with $E_\qq=\sqrt{(\epsilon_\qq+\mu)^2-\mu^2}$. 
Taking the derivative with respect to $\mu$, we get
$n=\mu/g_0+{1\over2}\sum_\qq(1-\epsilon_\qq/E_\qq)$. 

Now, using the relation between the bare repulsion $g$ and the boson scattering length $a_b$
given by
$M/4\pi a_b=1/g_0+\sum_\qq 1/2\epsilon_\qq$
we get 
\bqa
n & = & \mu M/4\pi a_b+ {1\over2}\sum_\qq\left[1-(\epsilon_\qq+\mu)/E_\qq\right]
\no \\
& & 
+ {\mu\over2}\sum_\qq\left(1/E_\qq-1/\epsilon_\qq\right),
\eqa 
where we have added and subtracted $\mu/2 E_\qq$ to isolate the cancellation
of divergences. The first integral is $-(1/3\pi^2)M^{3/2}\mu^{3/2}$ and 
the second one is $-M^{3/2}\mu^{3/2}/\pi^2$. So in all we get 
\beq
n = \mu M/4\pi a_b-(4/3\pi^2)M^{3/2}\mu^{3/2}
\eeq
We now solve this equation for $\mu(n)$ in powers of $(na_b^3)$.
To leading order $\mu=4\pi n a_b/M$ and to the next order in $a_b$ we get
\beq
\mu=4\pi n a_b/M[1+32/3 \pi^{-1/2}(na_b^3)^{1/2}]
\eeq
This is the correct equation of state for a Bogoliubov dilute Bose gas,
including the Lee-Yang correction.

We note that we \emph{cannot} identify the saddle point value of $\Phi_0^2$
in eq.~(\ref{mf}) with the condensate fraction, once quantum fluctuations
are taken into account. {This identification is usually made, together with the replacement $g_0 \rightarrow 4\pi a_b/M$. 
We note here that this identification makes (\ref{Omega Bosons}) divergent and is thus not well defined.}.  To find the condensate fraction we use
the expression for the $\qq \ne 0$ momentum distribution 
\beq
n(\qq) = {1 \over \beta}\sum_{iq_l} e^{iq_l0^+}\left( \A^{-1} \right)_{11}
\eeq
to derive the well known result
\beq
n(\qq) = {1 \over 2}\left({{\epsilon_\qq + \mu}\over E_\qq} - 1 \right), \ \ \ \ \ (\qq \ne 0).
\eeq
From the quantum depletion we can obtain the well known result for 
the condensate fraction using $N_0 = N - \sum_{\qq \ne 0} n(\qq) = N \left( 1- 8/3 \sqrt{(n a_b)^3/\pi} \right)$.



\begin{thebibliography}{99}
\bibitem{finite T}
At finite temperatures, however, the mean field theory only yields qualitatively correct results in the BCS limit, see Ref. \cite{sademelo}.
\bibitem{error}
We note some differences between the results here and
those in ref.~\cite{engelbrecht97}. First, there is a
a redefinition of variables in such a way that our 
$M_{11}$ and $M_{22}$ are interchanged. Second, we 
note two typographical sign errors in that reference: 
(a) The first line of eq.~(6) of ref.~\cite{engelbrecht97} defining $\MM_{12}$ 
should have an overall negative sign in front, which is corrected in eq.~(\ref{m12}) here.
(b) There is an incorrect minus sign in the definition of $Q$ in the paragraph below
eq.~(7) of ref.~\cite{engelbrecht97} which is corrected just above eq.~(\ref{speed of sound}) 
in this paper.
\bibitem{note on Gaussian 2}
{The condition under which this integration can be performed is that $X$, defined in (\ref{Xequation}), is positive; see \cite{note on Gaussian integration}.  This corresponds to the spectrum being gapped; if $X<0$ the system would be unstable.}
\end{thebibliography}
\end{document}